\documentstyle[epsfig]{elsart}

\newcommand{\be}{\begin{equation}}
\newcommand{\ee}{\end{equation}}

\newcommand{\bbf}{\bf}
\newcommand{\ssl}{\sl}

\newcommand{\Gamm}{\Upsilon}
\newcommand{\s}{{\bf s}}

\newcommand{\muo}{\mu_0}
\newcommand{\bea}{\begin{eqnarray}}
\newcommand{\eea}{\end{eqnarray}}

\renewcommand{\theequation}{\arabic{section}.\arabic{equation}}

\begin{document}

\begin{frontmatter}
\

\vspace{3 cm}

\title{Black hole pair creation in 
 de Sitter space: 
a complete one-loop analysis}

\author[MSV]{Mikhail S. Volkov\thanksref{DFG}} and
\author[AW]{Andreas Wipf}
\address{Institute for Theoretical Physics\\
Friedrich Schiller University of Jena\\
Max-Wien Platz 1, D-07743, Jena, Germany.}
\address[MSV]{e-mail: vol@tpi.uni-jena.de}
\address[AW]{e-mail: wipf@tpi.uni-jena.de}

\thanks[DFG]{Supported by the DFG}

\begin{abstract}
We present an exact one-loop calculation 
of the tunneling process in Euclidean quantum gravity
describing creation of black hole pairs  in a de Sitter universe. 
Such processes are mediated by $S^2\times S^2$ gravitational
instantons giving an imaginary contribution to the 
partition function.  
The required energy is provided by the expansion
of the universe. We utilize the thermal properties of de Sitter space
to describe the process as the decay of a metastable thermal
state. 
Within the  Euclidean path integral approach to gravity, we  
explicitly determine the spectra of the 
fluctuation operators, exactly calculate the one-loop
fluctuation determinants in the $\zeta$-function
regularization scheme, and check the agreement with 
the expected scaling behaviour. 
 Our results show a constant volume
density of created black holes at late times, and a very strong
suppression of the nucleation rate for 
small values of $\Lambda$. 
\end{abstract}

\end{frontmatter}

\newpage

\section{Introduction}
Instantons play an important role in 
flat space gauge field theory \cite{INSTANTONS}. 
Being stationary points of the Euclidean action,
they give the dominant contribution to the Euclidean
path integral thus accounting for a variety of 
important phenomena  in QCD-type theories. In addition, 
self-dual instantons
admit supersymmetric extensions, which makes them 
an important tool for verifying various duality 
conjectures  like the AdS/CFT correspondence \cite{Maldacena99}. 
More generally, the Euclidean approach has become 
the standard method of quantum field theories in 
flat space. 

Since the theory of gravity and Yang-Mills theory are somewhat 
similar, it is natural to study also gravitational instantons. 
An impressive amount of work
 has been done in this direction,
leading to a number of important discoveries. 
A thorough study of 
instanton solutions of the vacuum Einstein equations 
and also those with a $\Lambda$-term 
has been carried out \cite{Gibbons79LN,Eguchi80,GibbonsEQG}. 
These solutions dominate the path 
integral of Euclidean quantum gravity, leading to 
interesting phenomena like black hole nucleation and 
quantum creation of universes. 
Perhaps one of the most spectacular achievements of the 
Euclidean approach is the derivation of black 
hole entropy from the action of the Schwarzschild 
instanton \cite{Gibbons77}. In addition, gravitational 
instantons are used in the 
Kaluza-Klein reductions of string theory. 

Along with these very suggestive results, the difficulties of 
Euclidean quantum gravity have been revealed.
Apart from the usual problem of the 
non-renormalizability of gravity, which 
can probably be resolved only at the level of a more fundamental
theory like string theory,  the Euclidean approach presents 
other challenging problems. 
In field theories in flat space the correlation functions of field 
operators are holomorphic functions of the global
coordinates in a domain that includes negative imaginary values
of the time coordinate, $t=-i\tau$, where $\tau$ is real and positive
\cite{Streater64}. This allows one to perform the analysis in the 
Euclidean section and then analytically continue the functions
back to the Lorentzian sector to obtain the physical predictions. 
In curved space the theorems that would ensure the 
analyticity of any quantities arising in quantum gravity are not 
known.
As a result, even if Euclidean calculations
make sense, it is not in general clear how to relate their result to the 
Lorentzian physics.

This difficulty is 
most strikingly illustrated by the famous problem of the conformal
sector in Euclidean quantum gravity. 
If one tries evaluating the path integral over 
Riemannian metrics, then one discovers that it
diverges because the Euclidean gravitational action 
is not bounded from below and can be made arbitrarily large
and negative by conformal rescaling of the metric \cite{Gibbons78}. 
Such a result is actually expected, for if the integral did converge
(with some regularization), 
then one could give a well-defined meaning to the canonical 
ensemble of the quantum gravitational field. However, the 
possibility of having a black hole causes the canonical ensemble
to break down -- since the degeneracy of black hole states 
grows faster than the Boltzmann factor decreases. 
One can,  `improve' the Euclidean gravitational action by
analytically continuing the conformal modes, let us call them $h$,
via $h\to ih$, and this improves the 
convergence of the integral \cite{Gibbons78}. This shows
that if there is a well-defined Euclidean path integral for the 
gravitational field, then the relation to the Lorentzian sector 
is more complicated than just via $t\to -i\tau$. 

Unfortunately, it is unknown at present whether one can in the 
general case find a physically well-defined and convergent
path integral for the gravitational field. 
At the same time, the idea of 
constructing it is conceptually simple \cite{Schleich87}:
one should start from the Hamiltonian
path integral over the physical degrees of freedom
of the gravitational field. Such an integral 
certainly makes sense physically
and is well-convergent, since the Hamiltonian is positive --
at least in the asymptotically flat case. The Hamiltonian
approach is not covariant, but one can covariantize it
by changing the integration variables, which leads to a manifestly
covariant and convergent path integral for gravity. 
The main problem with this program is that in the general case 
it is unclear how to isolate the physical degrees of freedom of the 
gravitational field. For this reason, so far the program
has been carried out only for weak fields in the asymptotically
flat case \cite{Schleich87}. 
Remarkably, the result has been shown to exactly 
correspond to the the standard Euclidean path integral with the 
conformal modes being complex-rotated  via $h\to ih$. 
This lends support to the Euclidean
approach in gravity and allows one to hope that the difficulties
of the method can be consistently resolved;
(see, for example, \cite{Ambjorn00,Ambjorn00a} for the recent new 
developments within the lattice approach). 

One can adopt  the viewpoint that
Euclidean quantum gravity is a meaningful theory
within its range of applicability, at least at one-loop level,
by assuming that a consistent resolution of its difficulties exists. 
Then already in its present status the theory can be used
for calculating certain processes, most notably for describing
tunneling phenomena, in which case
the Euclidean amplitude directly determines the probability.
The analytic continuation to the Lorentzian 
sector in this case is not necessary, apart from when the issue of the 
interpretation of the corresponding gravitational instanton is 
considered. The important example of a tunneling process in 
quantum gravity is the creation of black holes in external fields. 
Black holes are created whenever the energy pumped into
the system is enough in order to make a pair of virtual 
black holes real \cite{Hawking96}. The energy can be provided by 
the heat bath \cite{Gross82,Kapusta84,Allen84}, by the background magnetic field 
\cite{Gibbons86,Garfinkle94,Dowker94,Dowker94a}, 
by the expansion of the universe 
\cite{Ginsparg83,Bousso96,Mann95}, by cosmic strings
\cite{Hawking95}, domain walls \cite{Caldwell96}, etc;
(see also \cite{Mellor89,Hawking94,Hawking95aa}). 
Besides, one can consider pair creation of extended
multidimensional 
objects like $p$-branes due to interaction with the 
background supergravity fields \cite{Dowker94aa}. 
In all these examples the process is mediated
by the corresponding gravitational instanton, and the 
semiclassical nucleation rate for a pair of objects 
on a given background is given by
\be                               \label{0}
\Gamma={ A}\exp\left\{-(I_{\rm obj}-I_{\rm bg})\right\}\, .
\ee
Here $I_{\rm obj}$ is the classical 
action of the gravitational instanton
mediating creation of the objects, $I_{\rm bg}$ is the
action of the background fields alone, and the 
prefactor ${ A}$ includes quantum corrections.  

In most cases the existing calculations of black hole
pair creation processes consider only the  
classical term in (\ref{0}). This is easily understood,
since loop calculations in quantum gravity for
non-trivial backgrounds are extremely complicated. 
To our knowledge, there is only one example of a 
next-to-leading-order
computation, which was undertaken in \cite{Gross82} 
by Gross, Perry, and Yaffe 
for the Schwarzschild instanton background.
The aim of the present paper is to consider one more example
of a complete one-loop computation in quantum gravity. 

The problem we are interested in is the quantum creation 
of black holes in de Sitter space. This problem was
considered  by Ginsparg and Perry \cite{Ginsparg83}, 
who identified the instanton responsible for this process,  
which is the $S^2\times S^2$ solution of the Euclidean Einstein
equations $R_{\mu\nu}=\Lambda g_{\mu\nu}$ for $\Lambda>0$. 
Ginsparg and Perry noticed that this solution has one 
negative mode in the physical sector, which renders the 
partition function complex, thus indicating the quasi-classical
instability of the system. 
This instability leads to 
spontaneous nucleation of black holes in the rapidly inflating
universe. This is the dominant instability of de Sitter space,
since classically the space is stable \cite{Ginsparg83}. 
The energy necessary for the nucleation is
provided by the $\Lambda$-term, which drives different parts
of the universe apart thereby drugging the members of a virtual
black hole pair away from each other. The typical radius
of the created black holes is $1/\sqrt{\Lambda}$, while the 
the nucleation rate is of the order of $\exp(-\pi/\Lambda G)$, where $G$
is Newton's constant. As a result, for $\Lambda G\sim 1$
when inflation is fast, the 
black holes are produced in abundance but they are small and
presumably almost immediately evaporate. Large black holes emerge
for $\Lambda G\ll 1$ when  inflation slows down, 
and these can probably exist for a long time, 
but the probability of their creation is exponentially
small.  This scenario was further studied in 
Refs.\cite{Bousso96,Bousso98,Elizalde99} 
(see also references in \cite{Bousso98}), where the generalization
to the charged case was considered and also the subsequent    
evolution of the created black holes was analyzed.  
However, the one-loop contribution so far has not been computed. 

A remarkable feature of the $S^2\times S^2$ instanton is its high
symmetry. In what follows, we shall utilize this symmetry
in order  to explicitly determine spectra of all relevant 
fluctuation operators in the problem. We shall use the $\zeta$-function 
regularization scheme in order to compute the one-loop
determinants, which will give us the partition function $Z[S^2\times S^2]$
for the small fluctuations around the $S^2\times S^2$ instanton. 
We shall then need to normalize this result. 
The normalization coefficients is 
$Z[S^4]$, the partition function for small fluctuations around the $S^4$
instanton, which is the Euclidean version of the de Sitter space. 
The one-loop quantization around the $S^4$ instanton was considered
by Gibbons and Perry \cite{Gibbons78a}, and by Christensen and Duff 
\cite{Christensen80}, but
unfortunately in none of these papers the analysis was 
completed. We shall therefore reconsider the problem 
by rederiving the spectra of fluctuations around $S^4$ 
and computing the determinants
within the $\zeta$-function scheme, thereby obtaining a closed
one-loop expression for $Z[S^4]$. 

In our treatment of the path integral we follow the approach of 
Gibbons and Perry \cite{Gibbons78a};
(see also \cite{Mazur89}). In order to 
have control over the results, we work in a 
one-parameter family of covariant gauges
and perform the Hodge decomposition of the fluctuations. 
These are then expanded with respect to the complete sets of 
basis harmonics, and the perturbative path integration measure
is defined as the square root of the determinant of the metric on the 
function space of fluctuations. To insure the convergence
of the integral over the conformal modes, which enter the action with the
wrong sign, we essentially follow the standard recipe $h\to ih$
\cite{Gibbons78}; (see also Ref.\cite{Mazur89}, where 
a slightly disguised form of the same prescription was advocated).
The subtle issue is that the 
conformal operator $\tilde{\Delta}_0=-3\nabla_\mu\nabla^\mu-4\Lambda$
has a finite number, ${\cal N}$,  of { negative} modes,
and these enter
the action with the correct sign from the very beginning. 
Our treatment of these special modes is different from that
by Hawking \cite{Hawking79}, who suggests that such modes should be
complex-rotated twice,  the partition function then acquiring the 
overall factor of $i^{\cal N}$. However, the presence of this 
factor in the partition function would lead to unsatisfactory results, 
and on these grounds we are led to 
not rotating the special conformal modes at all. 
 
The path integral is computed by
integrating over the Fourier expansion coefficients, which leads
to infinite products over the eigenvalues. The only conformal modes 
giving contribution to the result are the special negative modes 
discussed above.  We carefully analyze the resulting
products to make sure that all modes are taken into account
and that the dependence of the gauge-fixing
parameter cancels thereby indicating the correctness of the procedure. 
We  give a detailed consideration to the zero modes of the 
Faddeev-Popov operator, which arise due to the background isometries.
The integration over these modes requires a non-perturbative
extension of the path-integration measure, and we find such a
non-perturbative measure in the zero mode sector 
to be proportional to the Haar measure of the isometry group. 
Collecting all terms yields the  
partition function for small fluctuations around a background
instanton configuration in terms of infinite products over 
eigenvalues of the gauge-invariant operators. We then use the 
explicitly known spectra of fluctuations around the $S^2\times S^2$
and $S^4$ backgrounds in order to calculate the partition functions. 

The rest of the paper is organized as follows. In Sec.2 we present
our derivation of the black hole nucleation rate within the 
finite temperature approach. 
In Sec.3 the path integration procedure is considered. 
The spectra of small fluctuations around the $S^2\times S^2$
instanton are computed in Sec.4 via a direct solving of the 
differential equations in the eigenvalue problems. 
The spectra of the fluctuations around the $S^4$ instanton are rederived
in Sec.5 with the use of group theoretic arguments. 
The partition functions are computed in Sec.6, and Sec.7 
contains the final expression for the black hole nucleation rate
together with some remarks. We present a detailed analysis of the 
$\zeta$-functions in the Appendix.  We use units where 
$c=\hbar=k_{\rm B}=1$.

\section{Black hole nucleation rate}
In this section we shall derive the basic formula for the 
black hole nucleation rate in de Sitter space, whose 
different parts will be evaluated in the next sections.
The existing derivations of the nucleation rate 
\cite{Ginsparg83,Bousso96}  
recover only the classical factor in  (\ref{0}). 
In addition, it is not always clear to which volume the 
rate refers. We argue that our formula (\ref{rate}) gives
the nucleation probability  per Hubble volume 
and unit time as measured by a freely  
falling observer. 
The basic idea of our approach is to
utilize the relation between the inflation and  
thermal properties of de Sitter space. 
This will allow us to use the standard theory of decay
of metastable thermal states \cite{Langer67,Langer69,Affleck81}. 

Let us consider the partition function for the gravitational
field
\be                   \label{r1}
Z=\int D[g_{\mu\nu}]\,{\rm e}^{-I}\, ,
\ee
where the integral is taken over Riemannian metrics, and 
$I=I[g_{\mu\nu}]$ is the Euclidean action for gravity with a 
positive $\Lambda$ terms; see Eq.(\ref{1}) below.
The path integration procedure will be considered in detail in the next 
section. At present let us only
recall that in the semiclassical approximation the integral is
approximated by the sum over the classical extrema of the 
action $I$, that is 
\be            \label{r2}
Z\approx \sum_l Z_l\, .
\ee
Here $Z_l=Z[{\cal M}_l]$ is the partition function for the
small gravitational fluctuations around a background 
manifold ${\cal M}_l$ with 
a metric $g^l_{\mu\nu}$ subject to the Euclidean Einstein equations 
$R_{\mu\nu}=\Lambda g_{\mu\nu}$.
Schematically one has
\be                \label{r3}
Z\approx\sum_l\frac{\exp(-I_l)}{\sqrt{{\rm Det}\Delta_l}}\, ,
\ee
where $I_l$ is the classical action for the $l$-th 
extremum, and $\Delta_l$ is the operator for the 
small fluctuations around this background. 

The dominant contribution to the sum in (\ref{r3}) is given
by the $S^4$ instanton, which is the four-dimensional sphere
with the radius $\sqrt{3/\Lambda}$ and the standard metric. 
Since this is a maximally symmetry space, its action 
$I=-3\pi/\Lambda G$ is less than that of any other
instanton. Hence, 
\be                \label{r3a}
Z\approx Z[S^4]=
\frac{\exp(3\pi/\Lambda G)}{\sqrt{{\rm Det}\Delta}}\, .
\ee
On the other hand, 
the $S^4$ instanton describes the thermal properties
of de Sitter space \cite{Gibbons77,Gibbons77a}, since it 
can be obtained by an analytic continuation via 
$t\to \tau=it$
of the region of the de Sitter solution
\be                        \label{dS}
ds^2=-(1-\frac{\Lambda}{3} r^2)\,dt^2
+\frac{dr^2}{1-\frac{\Lambda}{3}r^2}
+r^2(d\vartheta^2+\sin^2\vartheta d\varphi^2)\, 
\ee
contained inside the 
event horizon, $r<\sqrt{3/\Lambda}$. 
Let us  call this region a Hubble region. Its
boundary, the horizon, has the area 
${\cal A}=12\pi/\Lambda$. The temperature associated with 
this horizon is $T=\frac{1}{2\pi}\sqrt{\frac{\Lambda}{3}}$, 
the entropy $S={\cal A}/4G=3\pi/\Lambda G$ 
and the free energy $F=-TS$.
The same values can be obtained by writing
the partition function for the $S^4$ instanton as
\be                \label{123}
Z[S^4]={\rm e}^{-\beta F}
\ee
with $\beta=1/T$. Indeed, since  
$S^4$ is periodic in all four coordinates,
any of them can be chosen to be the `imaginary time'. The 
corresponding period, $\beta=2\pi\sqrt{\frac{3}{\Lambda}}$, 
can be identified with the proper length of a geodesic on 
$S^4$, all of which are circles with the same length. 
This gives the correct de Sitter temperature. 
Comparing (\ref{123}) and  (\ref{r3a}) one obtains 
$\beta F=-{3\pi}/{\Lambda G}+\ldots$\,, the dots denoting the 
quantum corrections, and this again agrees
with the result for the de Sitter space. To recapitulate, the 
partition function of quantum gravity with 
$\Lambda>0$ is approximately 
\be
Z\approx {\rm e}^{-\beta F}\, ,
\ee
where $1/\beta$ is the de Sitter temperature and $F$ is the 
free energy in  the Hubble region. 

Let us now consider the contribution of the other instantons.
One has 
\be
Z\approx {\rm e}^{-\beta F}\left(1+\sum_{l}^{~~~~\prime}
\frac{Z[{\cal M}_l]}{Z[S^4]}\right)\, ,
\ee
where the prime indicates that  ${\cal M}_l\neq S^4$. 
Now, for $\Lambda G\ll 1$ all terms in the sum are exponentially
small and can safely be neglected as compared to the unity, 
if only they are real. If there are complex terms, then
they will give an exponentially small imaginary contribution.
The $S^2\times S^2$ instanton is distinguished
by the fact that its partition function is purely imaginary due to the 
negative mode in the physical sector \cite{Ginsparg83}.  
\begin{figure}[t]
\begin{minipage}[c]{13.5cm}
\centerline{\epsfysize=5 cm\epsffile{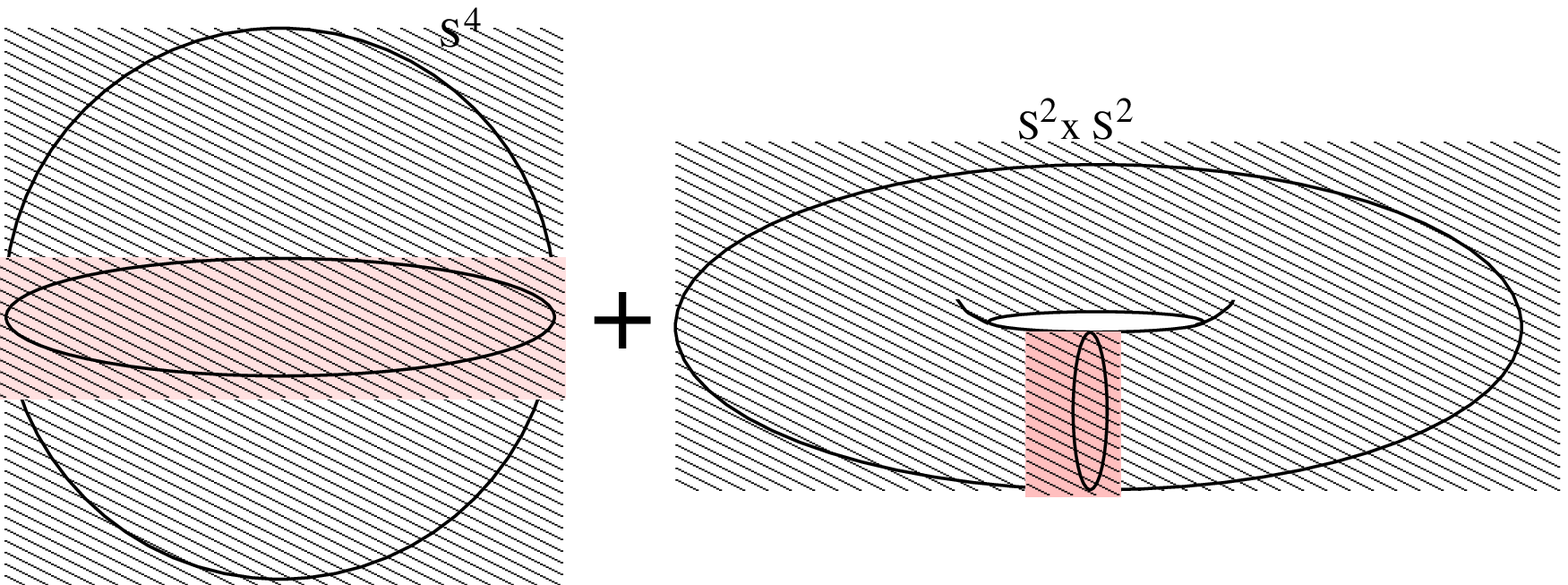}}
\caption
{
\label{INST}
{{\sl 
The leading contribution to the partition function comes
from the $S^4$ and $S^2\times S^2$ gravitational bubbles, 
the effect of the latter being purely imaginary. 
} 
}}
\end{minipage}
\end{figure}
This is the only solution  
for $\Lambda>0$ which is not a local minimum of the action
in the class of metrics with constant scalar curvature 
\cite{Gibbons79LN}. 
Hence (see Fig.\ref{INST}), 
\be
Z\approx {\rm e}^{-\beta F}\left(1+
\frac{Z[S^2\times S^2]}{Z[S^4]}\right)\approx
\exp\left(-\beta \left 
(F-\frac{Z[S^2\times S^2]}{\beta Z[S^4]}\right)\right)\, ,
\ee
where $Z[S^2\times S^2]$ is purely imaginary. 
As a result, the partition function can still be represented
as $Z\approx {\rm e}^{-\beta F}$, where the 
real part of $F$ is the free energy of the Hubble region, 
and the  exponentially small imaginary part is given by 
\be                      \label{ImF}
\Im (F)=-\frac{Z[S^2\times S^2]}{\beta Z[S^4]}\, .
\ee
It is natural to relate this imaginary quantity also to the free energy. 
We are therefore led to the conclusion that  the free
energy of the Hubble region has a small imaginary part, thus
indicating that the system is metastable. The decay of this
metastable state will lead to a spontaneous nucleation of a
black hole in the Hubble region, 
which can be inferred from the geometrical
properties of the $S^2\times S^2$ instanton. 

The $S^2\times S^2$ instanton can be obtained
via the analytic continuation of the Schwarzschild-de Sitter solution
\cite{Gibbons78b,Ginsparg83,Bousso96}
\be                           \label{SdS}
ds^2=-N\,dt^2
+\frac{dr^2}{N}
+r^2(d\vartheta^2+\sin^2\vartheta d\varphi^2)\, .
\ee
Here $N=1-\frac{2M}{r}-\frac{\Lambda}{3}r^2$, 
and for $9M^2\Lambda<1$ this function 
has roots at $r=r_{+}>0$ (black hole horizon) and
at $r=r_{++}>r_{+}$ (cosmological horizon). 
One has $N>0$ for $r_{+}<r<r_{++}$, and only this
portion of the solution can be analytically continued to the 
Euclidean sector via $t\to\tau =it$. 
The conical singularity at either of the 
horizons can be removed by a suitable identification 
of the imaginary time. 
However, since the two horizons have
different surface gravities, the second conical singularity 
will survive. The situation improves in the extreme limit, 
$r_{+}\to r_{++}\to\frac{1}{\sqrt{\Lambda}}$,
since the surface gravities are then the same and both conical
singularities can be removed at the same time. Although
one might think that the 
Euclidean region shrinks to zero when the two
horizons merge, this is not so. 
The limit $r_{+}\to r_{++}$ implies that 
$9M^2\Lambda=1-3\epsilon^2$ with $\epsilon\to 0$. 
One can introduce new coordinates  $\vartheta_1$ and 
$\varphi_1$ via
$\cos\vartheta_1=(\sqrt{\Lambda}r-1)/\epsilon+\epsilon/6$
and $\varphi_1=\sqrt{\Lambda}\,\epsilon\,\tau$. 
Passing to the new coordinates and taking the limit
$\epsilon\to 0$, the result is
\be                                            \label{SS}
ds^2=\left.\left.\frac{1}{\Lambda}\right(
d{\vartheta_1}^2+
\sin^2{\vartheta_1}\,d{\varphi_1}^2+
d{\vartheta}^2+
\sin^2\vartheta\,d{\varphi}^2\right)\, ,
\ee
and this $S^2\times S^2$ metric fulfills the Einstein equations.
Since the instanton field determines the initial value for the created
real time configuration, one concludes that the $S^2\times S^2$
instanton is responsible for the creation of a black hole in the 
Hubble region. This black hole fills the whole region, since
its size is equal to the radius of the cosmological horizon. 

\begin{figure}[t]
\begin{minipage}[c]{13.5cm}
\centerline{\epsfysize=6.5 cm\epsffile{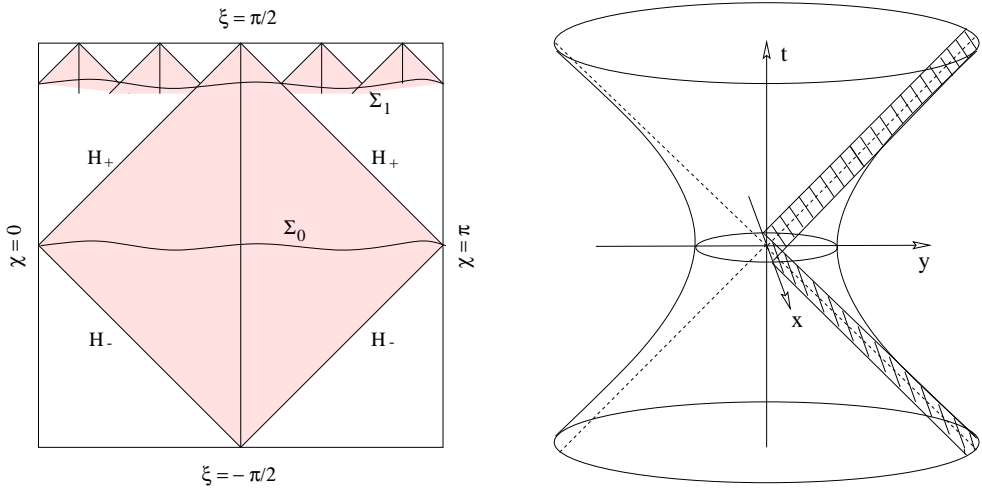}}
\caption
{
\label{CONFORMAL}
{{\sl {\rm Left:}
The conformal diagram of de Sitter space in coordinates
(\ref{dSglobal}). The trajectories 
$\chi$=const. are timelike geodesics. The diamond-shaped
region in the center is the Hubble region of the geodesic 
observer
at $\chi=\pi/2$. Although this  region 
completely covers the hypersurface $\Sigma_0$, at later times
one needs more  observers  
to cover the hypersurface $\Sigma_1$ with the interiors of their
horizons --
the Hubble regions proliferate.
{\rm Right:} The de Sitter hyperboloid in the embedding 
Minkowski space (with two dimensions suppressed). 
The Hubble region of the inertial observer
moving along the hyperbola $x=0$, $y>0$ 
is the portion
of the hyperboloid lying to the right from the two 
shaded strips. This corresponds to the interior region of 
the observer's static coordinate system. 
} 
}}
\end{minipage}
\end{figure}

It is well known that the region $r<\sqrt{{3}/{\Lambda}}$
of the static coordinate system in (\ref{dS})
covers only a small portion of the de Sitter hyperboloid
\cite{Schrodinger56}; (see Fig.\ref{CONFORMAL}). 
In order to cover the whole space,
one can introduce an infinite number of freely falling observers
and associate the interior of the static coordinate system 
with each of them. 
Hence, the spacetime contains infinitely many Hubble regions. 
It is also instructive to use global coordinates covering the 
whole de Sitter space,
\be                           \label{dSglobal}
ds^2=\left.\left.\frac{3}{\Lambda\cos^2\xi}\right(-d\xi^2
+d\chi^2
+\sin^2\chi\,(d\vartheta^2+\sin^2\vartheta d\varphi^2)\right)\, ,
\ee
where $\xi\in[-\pi/2,\pi/2]$ and $\chi\in[0,\pi]$. 
The trajectory of a freely falling observer is $\chi=\chi_0$  
(and also $\vartheta=\vartheta_0$, $\varphi=\varphi_0$),
and the domain of the associated static coordinate system,
the Hubble region, is the intersection of the interiors 
of the observer's past and future
horizons \cite{Hawking73}. Let $\Sigma$ be a spacelike
hypersurface, say $\xi=\xi_0$. If $\xi_0=0$ then $\Sigma$ is 
completely contained inside the Hubble region of a single
observer with $\chi=\pi/2$ (see Fig.\ref{CONFORMAL}). However, for 
late moments of time, $\xi\to \pi/2$, one needs more and more
independent observers in order to completely cover $\Sigma$
by the union of their Hubble regions. One can say that 
the Hubble regions proliferate with the expansion of the  universe. 

Since de Sitter space consists of infinitely many
Hubble regions, the black hole nucleation will lead to 
some of the regions being completely filled by
a black hole, but most of the regions will be empty. The 
number of the filled regions divided by the number of those
without a black hole is the 
probability for a black hole nucleation in one region. This 
is proportional to $\Im(F)$ in (\ref{ImF}). 

\begin{figure}[th]
\begin{minipage}[t]{13.5cm}
\centerline{\epsfysize=3.5 cm\epsffile{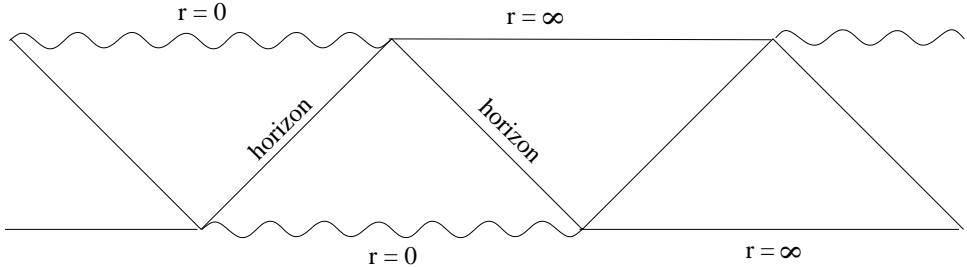}}
\caption
{
\label{SDS}
{{\sl 
The conformal digram for the extreme Schwarzschild-de Sitter
solution.} 
}}
\end{minipage}
\end{figure}

One can argue  that the black holes are actually created in pairs 
\cite{Hawking96,Hawking95aa}, where the two members of the 
pair are located at the antipodal points of the de Sitter 
hyperboloid. This can be inferred from the conformal diagram
of the Schwarzschild-de Sitter solution, 
which contains an infinite sequence of black hole singularities
and spacelike infinities; see Fig.\ref{SDS}. 
One can identify the asymptotically
de Sitter regions in the diagram related by a horizontal shift, 
and the spacetime will then 
consist of two black holes at antipodal points of the closed
universe. This agrees with the standard picture of particles in 
external fields being created in pairs. 

The surface gravity of the extreme Schwarzschild-de Sitter solution 
is finite when defined with respect to the suitably normalized Killing 
vector \cite{Bousso96}. This gives a non-zero value for the
 temperature
of the nucleated black holes, which 
can be read off also from the $S^2\times S^2$ metric: it is the 
inverse proper length of the equator of any of the two spheres, 
$T_{\rm BH}=\frac{\sqrt{\Lambda}}{2\pi}$. 
How can it be that 
this 
is different from the temperature of the heat bath,
which is the de Sitter space with
$T_{\rm dS}=\frac{1}{2\pi}\sqrt{\frac{\Lambda}{3}}$ ?
For example, in the hot Minkowski space 
the nucleated black holes have the same temperature as the 
heat bath \cite{Gross82}. 
However, the global 
structure of  de Sitter space is different
from that of Minkowski space. 
The fluctuations cannot {\sl absorb}  energy from 
{\sl and emit} energy into the 
whole of de Sitter space, but can only exchange  energy 
with  the Hubble region. 
Thus the energy exchange is restricted. As a result, the 
local temperature in the vicinity of a created defect may be
different from that of the heat bath, but reduces to the latter
in the asymptotic region far beyond the cosmological horizon. 

The relation of the imaginary part of the free energy to the 
rate of decay of a metastable thermal state $\Gamma$ 
was considered in \cite{Langer67,Langer69,Affleck81}. 
If the decay is only due to tunneling then $\Gamma=2\Im(F)$. 
Suppose that 
there is an additional possibility to classically jump over 
the potential barrier. In this case on top of the barrier
there is a classical saddle point configuration 
 whose real
time decay rate is determined by the saddle negative mode
$\omega_{-}$.  
At low temperatures the tunneling formula is then still correct,
while for $T>\frac{|\omega_{-}|}{2\pi}$ one has  
$\Gamma=\frac{|\omega_{-}|}{\pi T}\,\Im (F)$. 
In our problem the saddle point configuration also exists, the
$S^2\times S^2$ instanton, but its real time analog,
the Schwarzschild-de Sitter black hole, is stable. 
It seems therefore that there is no classical contribution to 
the process and the black hole nucleation is a purely quantum  
phenomenon.%
\footnote{We do not understand the 
classical interpretation of the Euclidean saddle point solution
suggested in \cite{Gross82}.  
The argument uses a family of non-normalizable deformations 
of the instanton, and the action is finite as long as they are `static'. 
However, if one considers a time evolution along such a family then 
the action will be infinite, which shows that the classical picture
does not apply. Even if one uses the classical formula for $\Gamma$ in this case, 
one arrives at the quantum result, since $|\omega_{-}|/T$=const.$\sim 1$.} 
[One can imagine that the effective potential barrier  
is infinitely high, such that a classical 
transition is forbidden, 
but at the same time so narrow that the tunneling rate is non-zero.]
As a result, the rate
of quasiclassical decay of the de Sitter space is given by 
$\Gamma=2\Im(F)$. Using Eq.(\ref{ImF}), 
\be                      \label{rate}
\Gamma=-2T\,\frac{Z[S^2\times S^2]}{Z[S^4]}\, .
\ee
Here $T=\frac{1}{2\pi}\sqrt{\frac{\Lambda}{3}}$
is the temperature of the de Sitter heat bath, which  was 
originally defined with respect to the 
analytically continued Killing vector $\frac{\partial}{\partial t}$.
Since $t$ is the proper time of the geodesic observer resting
at the origin of the static coordinate system (\ref{dS}), 
we conclude that the formula 
gives the probability of a black hole nucleation 
per Hubble volume and unit time of a freely falling observer. 

In order to use the formula (\ref{rate}), we should be able to 
compute the one-loop partition functions $Z[S^2\times S^2]$ 
and $Z[S^4]$. Now we shall calculate them 
within the path integral approach.

\section{The path integration procedure}

In this section we shall consider the path integral 
for fluctuations around an instanton solution of the 
Einstein equations $R_{\mu\nu}=\Lambda g_{\mu\nu}$ 
in the stationary phase approximation. 
We shall largely follow the approach of 
Gibbons and Perry  \cite{Gibbons78a}.

\setcounter{equation}{0}
\subsection{The second variation of the action}
Our starting point is  the action for the gravitational field
on a compact Riemannian manifold ${\cal M}$,
\be					\label{1}
I[g_{\mu\nu}]=-\frac{1}{16\pi G}\,\int_{\cal M} 
(R-2\Lambda) \sqrt{g}\, d^4x\, ,
\ee
whose extrema, $\delta I=0$, are determined
by the equations
\be					\label{2}
R_{\mu\nu}=\Lambda g_{\mu\nu}\, .
\ee
Let $g_{\mu\nu}$ be an arbitrary solution, and  
consider small fluctuations around it, 
$g_{\mu\nu}\to g_{\mu\nu}+h_{\mu\nu}$. 
The action expands as
\be					\label{3}
I[g_{\mu\nu}+h_{\mu\nu}]=I[g_{\mu\nu}]+
\delta^2 I+\ldots\, , 
\ee
where $\delta^2 I$ is quadratic in $h_{\mu\nu}$ 
and dots denote the higher order terms. 
One can express $\delta^2 I$ directly in terms
if $h_{\mu\nu}$. 
However, it is convenient to use first the 
standard decomposition of
$h_{\mu\nu}$,  
\be					\label{4}
h_{\mu\nu}=\phi_{\mu\nu}+\frac14\,h\, g_{\mu\nu}+
\nabla_\mu\xi_\nu+\nabla_\nu\xi_\mu-
\frac12\,g_{\mu\nu}\nabla_\sigma\xi^\sigma\, .
\ee
Here $\phi_{\mu\nu}$ is the transverse tracefree part,
$\nabla_{\mu}\phi^\mu_\nu=\phi^\mu_\mu=0$,
$h$ is the trace, and the piece due to $\xi_\mu$
is the longitudinal tracefree part. 
Under the gauge transformations (general diffeomorphisms)
generated by $\xi_\mu$ one has 
$
h_{\mu\nu}\to h_{\mu\nu}+\nabla_\mu\xi_\nu+\nabla_\nu\xi_\mu\, . 
$ 
The TT-tensor $\phi_{\mu\nu}$ is gauge-invariant, 
while the trace $h$ changes 
as $h\to h+2\nabla_\sigma\xi^\sigma$.
It follows that
\be					\label{5} 
\tilde{h}=h-2\nabla_\sigma\xi^\sigma
\ee
is gauge-invariant. For further references we note
that $\xi_\mu$ can in turn be decomposed into its
coexact part $\eta_\mu$, for which 
$\nabla_\mu\eta^\mu=0$, the exact part $\nabla_\mu\chi$,
and the harmonic piece $\xi_{\mu}^{\rm H}$,
\be					\label{6}
\xi_\mu=\eta_\mu+\nabla_\mu\chi+\xi_{\mu}^{\rm H}\, .
\ee
The number of square-integrable harmonic vectors
is a topological invariant, which is equal to the first
Betti number of the manifold ${\cal M}$. Since the latter is zero
if ${\cal M}$ is simply-connected, which is the case 
for $\Lambda>0$, we may safely ignore the
harmonic contribution in what follows.

With the decomposition (\ref{4}) the second
variation of the action in (\ref{3}) 
is expressed in terms
of the gauge-invariant quantities $\phi_{\mu\nu}$ and
$\tilde{h}$ alone,
\be					\label{7}
\delta^2 I=\frac12\,
\langle\phi^{\mu\nu},\Delta_2\phi_{\mu\nu}\rangle
-\frac{1}{16}\,\langle\tilde{h},\tilde{\Delta}_0
\tilde{h}\rangle\, .
\ee
Here and below we consider the following 
second order differential operators:
the operator for the TT-tensor fluctuations
\be					\label{8}
\Delta_2\phi_{\mu\nu}=-\nabla_\sigma\nabla^\sigma
\phi_{\mu\nu}
-2R_{\mu\alpha\nu\beta}\phi^{\alpha\beta}\, ,
\ee
the vector operator acting on coexact vectors $\eta_\mu$
\be					\label{9}
\Delta_1=-\nabla_\sigma\nabla^\sigma-
\Lambda\, ,
\ee
and the scalar operators for $h$, $\tilde{h}$, 
and $\chi$
\bea
\Delta_0&=&-\nabla_\sigma\nabla^\sigma\, , \nonumber \\
\tilde{\Delta}_0&=&3\Delta_0-4\Lambda\, , \nonumber \\
\tilde{\Delta}^{\gamma}_0&=&
\gamma\tilde{\Delta}_0-\Delta_0\,  ,     \label{10}
\eea
with $\gamma$ being a real parameter. 
Since for $\Lambda>0$ the manifold ${\cal M}$
is compact, 
these operators are (formally) self-adjoint with 
respect to the scalar product 
\be					\label{11}
\langle\phi_{\mu\nu},\phi^{\mu\nu}\rangle=
\frac{1}{32\pi G}\,
\int_{\cal M} 
\phi_{\mu\nu}\phi^{\mu\nu}\sqrt{g}\, d^4 x\, ;
\ee
similarly for vectors
$\langle\eta_{\mu},\eta^{\mu}\rangle$ 
and scalars $\langle\chi,\chi\rangle$. 

The action $\delta^2 I$ in (\ref{7}) contains
only the gauge-invariant amplitudes
$\phi_{\mu\nu}$ and $\tilde{h}$, while the 
dependence on the gauge degrees of freedom $\xi_\mu$
cancels. Pure gauge modes are thus zero modes of the 
action. Fixing of the gauge is therefore necessary 
in order to carry out the path integration. 
To fix the gauge we pass from the action $\delta^2 I$
to the gauge-fixed action 
\be					\label{12}
\delta^2 I_{gf}=\delta^2I+\delta^2I_g\, ,
\ee
where, following \cite{Gibbons78a}, we choose
the gauge-fixing terms as
\be					\label{13}
\delta^2 I_g=\gamma\left\langle
\nabla_\sigma h^\sigma_\rho-\frac{1}{\beta}\,
\nabla_\rho h,
\nabla^\alpha h_\alpha^\rho-\frac{1}{\beta}\,
\nabla^\rho h\right\rangle ,
\ee
with $\gamma$ and $\beta$ being real parameters. 
We shall shortly see that 
it is convenient to choose \cite{Gibbons78a}
\be					\label{14}
\beta=\frac{4\gamma}{\gamma+1}\, .
\ee
This choice, however, implies that $\delta^2 I_g$
does not vanish for $\gamma\to 0$. It is often 
convenient to set $\gamma=1$, in which case $\beta=2$.
However, we shall not fix the value of $\gamma$, since
this will provide us with a check of  the
gauge-invariance of our results. 

Using the decompositions 
(\ref{4}), (\ref{6}) the gauge-fixing term 
reads
\be					\label{15}
\delta^2 
I_g=\gamma\langle\eta_\mu,\Delta_1^2\eta^\mu\rangle
+\frac{1}{16\gamma}\langle
(\tilde{h}+2\tilde{\Delta}^\gamma_0\chi),
\Delta_0 (\tilde{h}+2\tilde{\Delta}^\gamma_0\chi)\rangle\, .
\ee
Adding this up with $\delta^2 I$ in 
(\ref{7}) one obtains the gauge-fixed action $\delta^2 
I_{gf}$. It is now convenient to pass 
from the gauge-invariant variable $\tilde{h}$ defined
in (\ref{5}) back to the trace $h$, since with the 
choice in (\ref{14}) the resulting action then becomes
diagonal:
\bea					\label{16}
\delta^2 I_{gf}&=&\frac12\,
\langle\phi^{\mu\nu},\Delta_2\phi_{\mu\nu}\rangle
+\gamma\langle\eta_\mu,\Delta_1^2\eta^\mu\rangle \\
&+&\frac14\,\langle\chi,\Delta_0\tilde{\Delta}_0
\tilde{\Delta}_0^\gamma\chi\rangle
-\frac{1}{16\gamma}\,\langle h,\tilde{\Delta}_0
h\rangle\, . \nonumber 
\eea
This action generically has no zero modes,
but it depends on the arbitrary parameter $\gamma$,
which reflects the freedom of choice of gauge-fixing.
In order to cancel this dependency, the compensating
ghost term is needed.

\subsection{The mode decomposition of the action} 
We wish to calculate the path integral
\be					\label{17}
Z[g_{\mu\nu}]={\rm e}^{-I}\int D[h_{\mu\nu}]\,{\cal D}_{\rm FP}
\exp\left(-\delta^2 I_{gf}\right),
\ee
where 
$I=I[g_{\mu\nu}]$ is the classical action, and 
the Faddeev-Popov factor is obtained from
\be					\label{18}
1={\cal D}_{\rm FP}
\int D[\xi_\mu]\,
\exp\left(-\delta^2 I_{g}\right).
\ee
In order to perform the path integration, 
we introduce the 
eigenmodes associated with the operators
$\Delta_2$, $\Delta_1$ and $\Delta_0$:
\bea
\Delta_2\,\phi^{(k)}_{\mu\nu}&=&
\varepsilon_k\,\phi^{(k)}_{\mu\nu}, \nonumber \\
\Delta_1\,\eta_{\mu}^{(s)}&=&
\sigma_s\,\eta_{\mu}^{(s)}, \nonumber \\
\Delta_0\,\alpha^{(p)}&=&
\lambda_p\,\alpha^{(p)}.           \label{19}
\eea
Throughout this paper we shall denote  
the eigenvalues and eigenfunctions of the tensor
operator $\Delta_2$
by $\varepsilon_k$ and $\phi^{(k)}_{\mu\nu}$, 
and those for the vector operator $\Delta_1$ by $\sigma_s$
and $\eta_{\mu}^{(s)}$, respectively. 
[Later we shall use the symbol $s$ also for the argument
of the $\zeta$-functions, and this will not lead to any
confusion].
Eigenvalues of the scalar operator will be denoted by 
$\lambda_p$, and it will be convenient to split the 
set $\{\lambda_p\}$ into three subsets,
$\{\lambda_p\}=\{\lambda_0,\lambda_i,\lambda_n\}$, 
where $\lambda_0=0$, $\lambda_i=\frac43\Lambda$, and
$\lambda_n>\frac43\Lambda$; see Eqs.(\ref{25})--(\ref{27}) below.
Accordingly, the set of the scalar eigenfunctions will be split as
$\{\alpha^{(p)}\}=\{\alpha^{(0)},\alpha^{(i)},\alpha^{(n)}\}$.

Since the manifold is compact, we choose the modes to be
orthonormal. 
This allows us to expand all fields in the problem
as
\be					\label{20}
\phi_{\mu\nu}=\sum_k C_k^\phi \phi_{\mu\nu}^{(k)}\, ,
\ \ \ \eta_{\mu}=\sum_s C_s^\eta \eta_{\mu}^{(s)}\, ,
\ee
and 
\be                                   \label{20a}   
\chi=\sum_p C_p^\chi \alpha^{(p)}\, ,\ \ \
h=\sum_p C_p^h \alpha^{(p)}\, ,\ \ \ 
\tilde{h}=\sum_p C_p^{\tilde{h}} 
\alpha^{(p)}\,.
\ee
As a result, the action decomposes into the sum
over modes, and the path integral
reduces to integrals 
over the Fourier coefficients. 

{\bf a) Vector and tensor modes.--}
Let us consider the mode 
decomposition for the gauge-fixed action in (\ref{16}). 
This action 
is the sum of four terms. For the first 
two terms we obtain 
\bea				    \label{21}
\frac12\,
\langle\phi^{\mu\nu},\Delta_2\phi_{\mu\nu}\rangle&=&
\frac12\,\sum_k\varepsilon_k\,
(C_k^\phi)^2\, , \\
\gamma\,\langle\eta_\mu,\Delta_1^2\eta^\mu\rangle 
&=&\gamma\,\sum_s(\sigma_s)^2
(C_s^\eta)^2 \, .     \label{22}
\eea
These quadratic forms should be
positive definite, since otherwise the integrals
over the coefficients $C$ would be ill-defined.   
We can see that 
the quadratic form in (\ref{22}) for the
vector modes is indeed non-negative definite. 
Next, the expression in 
(\ref{21}) for the gauge-invariant tensor modes is
positive-definite if all eigenvalues $\varepsilon_k$
are positive. If there is a negative
eigenvalue, $\varepsilon_{-}<0$, 
as in the case of the $S^2\times S^2$ instanton background, 
then it is physically significant. The integration over
$C^\phi_{-}$ is performed with the complex
contour rotation, which renders the partition function
imaginary thus indicating the 
quasiclassical instability of the system.

Let us consider now the contribution of the longitudinal
vector piece to the action (\ref{16}).  We obtain 
\be				     \label{24a}
\frac14\,\langle\chi,\Delta_0\tilde{\Delta}_0
\tilde{\Delta}_0^\gamma\chi\rangle=\frac14\,
\sum_p
\lambda_p\tilde{\lambda}_p
\tilde{\lambda}^{\gamma}_{p}\,
(C_p^\chi)^2 \, ,
\ee
where
$\tilde{\lambda}_p
=3{\lambda}_p-4\Lambda$ and 
$\tilde{\lambda}_{p}^{\gamma}=
\gamma\tilde{\lambda}_p-
{\lambda}_p$
are the eigenvalues of $\tilde{\Delta}_0$ and
$\tilde{\Delta}_0^\gamma$, respectively. 
We note that while $\lambda_p\geq 0$,
the $\tilde{\lambda}_p$ and 
$\tilde{\lambda}_{p}^{\gamma}$ can be negative
and should therefore be treated carefully.    
Let us split the scalar modes 
into three groups according to the sign of 
$\tilde{\lambda}_p$:
\bea                                   \label{25}
&&\Delta_0\,\alpha^{(0)}=0, \ \ \ \ \ \ \ \ 
\ \ \ \Rightarrow
\ \  \tilde{\lambda}_0=-\frac43\,\Lambda\, , \\
&&\Delta_0\,\alpha^{(i)}
=\frac43\,\Lambda\,\alpha^{(i)},
\ \ \ \Rightarrow
\ \  \tilde{\lambda}_i=0\, ,\label{26} \\
&&\Delta_0\,\alpha^{(n)}
=\lambda_n\,\alpha^{(n)},
\ \ \ \Rightarrow
\ \  \tilde{\lambda}_n>0\, .\label{27}
 \eea
First we consider  the constant mode $\alpha^{(0)}$ 
in (\ref{25}). This exists for any background, and for 
compact manifolds without boundary this is the only
normalizable zero mode of $\Delta_0$. 
Since this mode is annihilated by ${\Delta}_0$, it  
does not contribute to the sum in (\ref{24a}).  

Consider now the scalar modes with the eigenvalue 
$4\Lambda/3$ in (\ref{26}). In view of the Lichnerowicz-Obata
theorem \cite{Yano70}, the lowest non-trivial eigenvalue of 
$\Delta_0$ for $\Lambda>0$ 
is bounded from below by $4\Lambda/3$,
and the equality is attained if only the background is $S^4$. 
Hence the modes in (\ref{26}) exist only for the $S^4$ instanton,
and there can be 
no modes `in between' (\ref{25}) and (\ref{26}). 
In the $S^4$ case there are five scalar modes
with the eigenvalue $4\Lambda/3$, and their
gradients are the five conformal Killing vectors that do not correspond to 
infinitesimal isometries. If $R_{\mu\nu}=\Lambda g_{\mu\nu}$, 
a theorem of Yano an Nagano \cite{Yano70} states that such vectors 
exist only in the $S^4$ case. Let us  call  these five scalar modes 
`conformal Killing modes'. Notice that these also 
do not contribute to the sum in (\ref{24a}). 

To recapitulate, the lowest lying modes in the scalar spectrum 
are the constant conformal mode in (\ref{25}), which exists for any
background, and also
 5 `conformal Killing modes' in (\ref{26}) which exist only
for the $S^4$ instanton and generate the conformal isometries. 
As we shall see, these 1+5 lowest lying modes are physically distinguished,
since they are the only scalar modes contributing 
to the partition function. However, they do not
enter the sum in (\ref{24a}).

For the remaining infinite number of scalar modes in (\ref{27}) 
(these are labeled by $n$)
the eigenvalues $\lambda_n$
and $\tilde{\lambda}_n$ are positive, 
and it is not difficult to see that all the 
$\tilde{\lambda}_{n}^{\gamma}$'s are also positive,
provided that the gauge parameter $\gamma$ is positive
and large enough.
To recapitulate, the contribution
of the longitudinal vector modes to the action
is given by
\be				     \label{24}
\frac14\,\langle\chi,\Delta_0\tilde{\Delta}_0
\tilde{\Delta}_0^\gamma\chi\rangle=\frac14\,
\sum_n
\lambda_n\tilde{\lambda}_n
\tilde{\lambda}^{\gamma}_{n}\,
(C_n^\chi)^2 \, ,
\ee
which is positive definite. We shall  see that all  modes
contributing to this sum 
are unphysical and cancel from the path integral.

{\bf b) Conformal modes.--}
We now turn to the last term in the gauge-fixed
action (\ref{16}). Using (\ref{25})--(\ref{27})
we obtain 
\be                                   \label{28}
-\frac{1}{16\gamma}\,\langle h,\tilde{\Delta}_0
h\rangle=\frac{\Lambda}{4}\, 
(C^h_0)^2+
\frac{\Lambda}{12\gamma}\,\sum_i 
(C^h_i)^2   
-\frac{1}{16\gamma}\,\sum_n \,
\tilde{\lambda}_n^{\gamma}\,
(C^h_n)^2\, .
\ee
The expression on the right has a finite
number of positive terms, corresponding to the distinguished
lowest lying modes, and infinitely many 
negative ones. 
As a result, an increase in the coefficients $C^h_n$
makes it arbitrarily large and negative, 
thus rendering the path integral divergent. 
This represents the well-known problem of 
conformal modes in Euclidean quantum gravity
\cite{Gibbons78}. 
A complete solution of this problem is lacking
at present, but the origin of the trouble
seems to be understood \cite{Schleich87}. 
In brief,  
the problem is not related to any 
defects of the theory itself, but arises as a result
of the bad choice of the path integral.  
If one starts from the fundamental
Hamiltonian path integral over the physical
degrees of freedom of the gravitational field,  
then one does not encounter this problem. 
The Hamiltonian 
path integral, however, is non-covariant
and difficult to work with.  
One can `covariantize' it
by adding gauge degrees of freedom, and this
leads to the Euclidean path integral
described above, up to the 
important replacement \cite{Gibbons78}
\be					\label{29}
h\to ih\, .
\ee
The effect of this is to  
change the overall sign in (\ref{28}), such that 
the infinite number of negative modes
become positive. 
Unfortunately, such a consistent 
derivation of the path integral
has only been carried out for weak gravitational fields
\cite{Schleich87} (and for $\Lambda=0$), 
since otherwise it is unclear
how to choose the physical degrees of freedom. 
Nevertheless, the rule (\ref{29}) is often 
used also in the general case \cite{Gibbons78}, 
and it leads to the cancellation of the unphysical conformal modes. 
However, some subtle issues can arise. 

For $\Lambda>0$ 
the expression in (\ref{28}) contains, 
apart from infinitely many negative terms,
also a finite number of positive ones, which are due
to the distinguished lowest lying scalar modes. 
If we apply the rule (\ref{29}) and change
the overall sign of the scalar mode action,
 then the negative modes will become 
positive, but the positive ones will become negative.
As a result,
the path integral will still be divergent. 
It was therefore suggested by Hawking 
that the contour for these extra negative 
modes should be
rotated back, the partition function then acquiring 
the factor $i^{\cal N}$, where ${\cal N}$ is the 
number of such modes \cite{Hawking79}. As we know,
${\cal N}=6$ for the $S^4$ instanton, and ${\cal N}=1$
for any other solution of $R_{\mu\nu}=\Lambda g_{\mu\nu}$
with $\Lambda>0$.   

Unfortunately, this prescription to rotate the contour
twice leads in some cases to physically meaningless results;
the examples will be given in a moment. 
We suggest therefore a slightly different scheme: 
not to touch the positive modes in (\ref{28})
at all and to change 
 the sign only for the negative modes. 
The whole expression then becomes 
\be                                   \label{30}
-\frac{1}{16\gamma}\,\langle h,\tilde{\Delta}_0
h\rangle=\frac{\Lambda}{4}\, 
(C^h_0)^2+
\frac{\Lambda}{12\gamma}\,\sum_i 
(C^h_i)^2  
+\frac{1}{16\gamma}\,\sum_n \,
\tilde{\lambda}_n^{\gamma}\,
(C^h_n)^2\, .
\ee 
We make no attempt to rigorously justify such a 
rule. We note, however, that it is essentially
equivalent to the standard recipe (\ref{29}) --
up to a finite number of modes which we handle
differently as compared to Hawking's prescription. 
We shall now comment on this difference. 

When compared to Hawking's recipe
\cite{Hawking79}, 
the ultimate effect of our prescription 
is to remove the
factor $i^{\cal N}$ from the partition function. 
We are unaware of any examples where 
it would be necessary to insist on
this factor being present in the final result. 
On the contrary, the examples are
in favour of the factor being absent.  
For  the $S^4$ instanton one has
${\cal N}=6$, such that $i^{\cal N}=-1$, 
and this would render
the partition function 
for hot gravitons in a de Sitter universe
negative, which would be physically meaningless. 
Next, 
for the $S^2\times S^2$ instanton,
which already has one negative mode in the spin-2
sector, one has ${\cal N}=1$. As a result,  
the factor $i^{\cal N}$ would make the
partition function real instead of being
imaginary, and there would be no  black hole
pair creation ! 

These arguments suggest that 
Hawking's rule should be somehow modified,
and we therefore put forward  the prescription
 (\ref{30}).  
Let us also note that our rule leads to gauge
invariant results -- the dependence on the 
gauge parameter $\gamma$ cancels after the integration.  
Finally we note that the lowest lying scalar modes are
physically distinguished, and since they are positive, they 
should be treated similarly
to the physical tensor modes.

To recapitulate, the mode expansion of the 
gauge-fixed action $\delta^2 I_{gf}$ is given by the sum
of (\ref{21}), (\ref{22}), (\ref{24}),
and (\ref{30}): 
\bea				    \label{30a}
\delta^2 I_{gf}=
\frac12\,\sum_k\varepsilon_k\,
(C_k^\phi)^2+\gamma\,\sum_s(\sigma_s)^2
(C_s^\eta)^2+
\frac14\,
\sum_n
\lambda_n\tilde{\lambda}_n
\tilde{\lambda}^{\gamma}_{n}\,
(C_n^\chi)^2 \nonumber \\
+\frac{\Lambda}{4}\,
(C^h_0)^2+
\frac{\Lambda}{12\gamma}\,\sum_i 
(C^h_i)^2  
+\frac{1}{16\gamma}\,\sum_n \,
\tilde{\lambda}_n^{\gamma}\,
(C^h_n)^2\, .
\eea

In a similar way we obtain the following mode
expansion for the gauge-fixing term $\delta^2 I_{g}$
in (\ref{15}):
\bea				       \label{31}
\delta^2 I_{g}
&=&\gamma\,\sum_s(\sigma_s)^2
(C_s^\eta)^2 
+\frac{16}{27\gamma}\,\Lambda^3\sum_i 
\left(C_i^\chi
-\frac{3}{8\Lambda}C^{\tilde{h}}_{i}\right)^2\nonumber \\
&+&\left.\left.\frac{1}{16\gamma}\,\sum_n \lambda_n\, 
\right(2\tilde{\lambda}^\gamma_n\, C^\chi_n
+C^{\tilde{h}}_n\right)^2\, . 
\eea
This  expression is non-negative definite. 

\subsection{The path integration measure} 
In order to compute the path integrals in
(\ref{17}),(\ref{18}) we still need 
to define the path-integration measure. 
The perturbative measure is defined as 
the square root of the determinant of the 
metric on the function space
of fluctuations:
\be                                    \label{32}
D[h_{\mu\nu}]\sim\sqrt{{\rm Det}(\langle dh_{\mu\nu}, 
dh^{\mu\nu}\rangle})\, ,\ \ \ \ \
D[\xi_{\mu}]\sim\sqrt{{\rm Det}(\langle d\xi_{\mu}, 
d\xi^{\mu}\rangle})\, .\ \ 
\ee
Here it is assumed that the fluctuations are 
Fourier-expanded
and the differentials refer to the Fourier 
coefficients, while the meaning of the
proportionality sign will become clear shortly. 
Let us first consider $D[\xi_\mu]$. 
It follows from 
(\ref{4}),(\ref{6}) that 
\bea
\langle h_{\mu\nu},h^{\mu\nu}\rangle&=&
\langle\phi_{\mu\nu},\phi^{\mu\nu}\rangle
+2\langle\eta_\mu,\Delta_1\eta^\mu\rangle
+\langle\chi,\Delta_0\tilde{\Delta}_0\chi\rangle
+\frac14\,\langle h,h\rangle\, ,\nonumber \\
\langle \xi_{\mu},\xi^{\mu}\rangle&=&
\langle\eta_\mu,\eta^\mu\rangle
+\langle\chi,\Delta_0\chi\rangle\, .   \label{33}
\eea
Expanding the fields on the right according to 
(\ref{20}),(\ref{20a}) and differentiating with respect
to the Fourier 
coefficients we obtain the metric
for the vector fluctuations
\bea
\langle d\xi_{\mu},d\xi^{\mu}\rangle=
\langle d\eta_\mu,d\eta^\mu\rangle
+\langle d\chi,\Delta_0 d\chi\rangle 
=
\sum_s 
(dC_s^\eta)^2\, 
+\sum_p^{\ \ \ \prime} \lambda_p\,
(dC_p^\chi)^2 \, ,     \label{34}
\eea
which yields
\be                                    \label{35}
\sqrt{{\rm Det}(\langle d\xi_{\mu},d\xi^{\mu})
\rangle}=\left(
\prod_s 
dC_s^\eta \right)\left(
\prod_p^{\ \ \ \prime} \sqrt{\lambda_p}\,
dC_p^\chi \right) .
\ee
Here the prime indicates that terms with 
$\lambda_p=0$ do not contribute to the sum in
(\ref{34}), 
and should therefore be omitted in the 
product in (\ref{35}).  
To obtain the measure $D[\xi_\mu]$ we endow 
each term in the products in (\ref{35})
with the weight factor 
$\mu_o^2/\sqrt{\pi}$:
\be                                    \label{36}
D[\xi_\mu]=\left(
\prod_s
\frac{\mu_o^2}{\sqrt{\pi}}\,
\,
{dC_s^\eta} \right)\left(
\prod_i
\frac{\mu_o^2}{\sqrt{\pi}}\,
\,\sqrt{\frac{4\Lambda}{3}}\,
{dC_i^\chi} \right)\left(
\prod_n
\frac{\mu_o^2}{\sqrt{\pi}}\,
\,\sqrt{\lambda_n}\,
{dC_n^\chi} \right).
\ee
Such a normalization implies that 
\be			                \label{37}
1=\int D[\xi_\mu]\exp\left(-\mu_o^4\, 
\langle \xi_{\mu},\xi^{\mu}\rangle\right)\,.
\ee
Here $\mu_o$ is a 
parameter with the dimension of an inverse length. 
In a similar way we obtain the measure
$D[h_{\mu\nu}]$, which is normalized as 
\be			                \label{38}
1=\int D[h_{\mu\nu}]\exp\left(-\frac{\mu_o^2}{2}\, 
\langle h_{\mu\nu},h^{\mu\nu}\rangle\right)\, ;
\ee
we shall shortly comment on the relative normalization
of $D[h_{\mu\nu}]$ and $D[\xi_\mu]$.
The result is 
\bea                                    \label{39}
D[h_{\mu\nu}]&=&\left(
\prod_k\frac{\mu_o}{\sqrt{2\pi}}\,
\,dC_k^\phi \right)\left(
\prod_s^{\ \ \ \prime}
\frac{\mu_o}{\sqrt{2\pi}}\,
\,\sqrt{2\sigma_s}\, dC_s^\eta\right)\left(         
\prod_n
\frac{\mu_o}{\sqrt{2\pi}}\,
\,\sqrt{\lambda_n\tilde{\lambda}_n}\,
dC_n^\chi\right) \nonumber  \\
&\times&\,\left(
\frac{\mu_o}{\sqrt{2\pi}}
\,\frac12\,
dC_0^h\right) \left(
\prod_i
\frac{\mu_o}{\sqrt{2\pi}}
\,\frac12\,
dC_i^h\right)\left(
\prod_n
\frac{\mu_o}{\sqrt{2\pi}}\,
\,\frac12\,
dC_n^h\right) .
\eea
Here the prime indicates that the zero modes of the 
vector fluctuation operator 
do not contribute to the product. 
Notice, however, that these modes do contribute 
to the measure $D[\xi_\mu]$. 

The following remarks are in order. We use units where
all fields and parameters 
have dimensions of different powers of a length scale $l$. 
One has $[1/G]=[\Lambda]=[\mu_o^2]=[l^{-2}]$.
Eigenvalues of all  
fluctuation operators have the dimension $[l^{-2}]$.
The coordinates $x^\mu$ are dimensionless, 
while $[g_{\mu\nu}]=[h_{\mu\nu}]=[l^2]$. For the 
vectors, $[\eta_\mu]=[\xi_\mu]=[l^2]$,
and for the scalars $[h]=[l^0]$ and  
$[\chi]=[l^2]. $ 
We assume that
the scalar, vector and tensor eigenfunctions in (\ref{19})  
are orthonormal with respect to the 
scalar product in (\ref{11}). As a result, 
the dimensions of the eigenfunctions are
$[\phi^{(k)}_{\mu\nu}]=[l]$, 
$[\eta^{(s)}_\mu]=[l^0]$, $[\alpha^{(p)}]=[l^{-1}]$, 
which gives for  the Fourier coefficients in (\ref{20}),(\ref{20a}) 
$[C^\phi]=[C^h]=[l]$, $[C^\eta]=[l^2]$, and
$[C^\chi]=[l^3]$. 

The normalization of $D[h_{\mu\nu}]$ can be arbitrary,
which is reflected in the presence of the arbitrary parameter 
$\mu_o$ in the above formulas. However, the relative 
normalization of $D[h_{\mu\nu}]$ and $D[\xi_\mu]$, which is 
defined by Eqs.(\ref{37}) and (\ref{38}) is fixed by 
gauge invariance. Had we chosen instead a 
different relative normalization, say dividing 
each mode in (\ref{36}) by 2, then the path integral 
would acquire a factor of $2^{\tilde{N}^\gamma_0}$, where 
$\tilde{N}^\gamma_0$ is the `number of eigenvalues' of the 
non-gauge-invariant operator $\tilde{\Delta}^\gamma_0$.  
[The issue of relative normalization   
of the fluctuation and Faddeev-Popov determinants 
seldom arises, since in most cases the 
absolute value of the path integral is irrelevant].

\subsection{Computation of the path integral}
Now we are ready to 
compute the path integrals
in (\ref{17}),(\ref{18}). Let us illustrate
the procedure on the example of  
Eq.(\ref{18}), which determines the Faddeev-Popov
factor ${\cal D}_{FP}$. 
Using $\delta^2 I_{g}$ from Eq.(\ref{31}) and 
the measure $D[\xi_\mu]$ from (\ref{36})
we obtain 
\bea                                 \label{40}
&&({\cal D}_{FP})^{-1}=
\prod_s\int
\frac{\mu_o^2}{\sqrt{\pi}}\,
\,
{dC_s^\eta} \,
\exp\left(-\gamma(\sigma_s)^2
(C_s^\eta)^2 \right)  \\
&&\times
\prod_i\int
\frac{\mu_o^2}{\sqrt{\pi}}\,
\,\sqrt{\frac{4\Lambda}{3}}\,
{dC_i^\chi} \,\exp\left(-
\frac{16}{27\gamma}\,\Lambda^3
\left(C_i^\chi
-\frac{3}{8\Lambda}C^{\tilde{h}}_{i}\right)^2\right)
\nonumber \\
&&\times\prod_n\int
\frac{\mu_o^2}{\sqrt{\pi}}\,
\,\sqrt{\lambda_n}\,
{dC_n^\chi} \, 
\exp\left(-
\left.\left.\frac{1}{16\gamma}\,\sum_n \lambda_n\, 
\right(2\tilde{\lambda}^\gamma_n\, C^\chi_n
+C^{\tilde{h}}_n\right)^2\right), \nonumber
\eea
which gives
\be                                       \label{41}
({\cal D}_{FP})^{-1}=\Omega_1\left(
\prod_s^{\ \ \ \prime}
\frac{\mu_o^2}{\sqrt{\gamma}\sigma_s}\right)\left(
\prod_i
\frac{3\sqrt{\gamma}\mu_o^2}{2\Lambda}\right)\left(
\prod_n\frac{2\sqrt{\gamma}\mu_o^2}
{\tilde{\lambda}_n^\gamma}\right) .
\ee

\subsubsection{Zero modes}
The factor $\Omega_1$ in (\ref{41}) arises due to the 
gauge zero modes, for which
$\sigma_s\equiv\sigma_0=0$ and the 
integral is non-Gaussian:
\be                                 \label{42}
\Omega_1=\int\,
\prod_j
\frac{\mu_o^2}{\sqrt{\pi}}\,
\, {dC_{0j}^\eta} \, ,
\ee 
with the product taken over all  such modes. 
The existence of zero modes of the Faddeev-Popov operator
indicates that the gauge is not completely fixed. 
This can be related to the global aspects of gauge
fixing procedure known as the Gribov ambiguity.
However, Gribov's problem is usually not the issue in the 
perturbative calculations, where
zero modes arise rather due to  background symmetries. 
This will be the case in our analysis below. Specifically, the isometries
of the background manifold ${\cal M}$ form a subgroup ${\cal H}$
of the full diffeomorphism group.  Sometimes 
${\cal H}$ is called the stability group;  
for  the $S^4$ and $S^2\times S^2$ backgrounds
${\cal H}$ is SO(5) and SO(3)$\times$SO(3), respectively. 
Since the isometries do not change $h_{\mu\nu}$ 
(in the linearized approximation), their generators, 
which are the Killing vectors $K^\mu$, are zero modes 
of the Faddeev-Popov operator. 

We therefore conclude that the integration in (\ref{42}) 
is actually performed over the stability group ${\cal H}$. 
Since the latter is compact in the cases under consideration, 
the integral is finite. In order to actually compute
the integral, some further analysis is necessary, in which
we shall adopt the approach of Osborn \cite{Osborn81}. 
First of all, let us recall that all eigenmodes in our analysis have unit
norm. If we now rescale the zero modes  such that 
the Killing vectors 
$K_j\equiv K_j^\mu\frac{\partial}{\partial x^\mu}$ become
dimensionless (remember that the coordinates $x^\mu$ are also 
dimensionless), then the expression in Eq.(\ref{42}) reads
\be                                 \label{42aa}
\Omega_1=\int\,
\prod_j
\frac{\mu_o^2}{\sqrt{\pi}}\,||K_j||
\, {dC_j} \, ,
\ee 
where now $[||K_j||]=[l^2]$ and $[C_j]=[l^0]$. 
For small values of the parameters $C_j$ they can 
be regarded as coordinates on the 
group manifold ${\cal H}$ in the vicinity of the unit element. 
Since 
${\cal H}$ acts on ${\cal M}$ via 
${x}^\mu\to {x}^\mu(C_j)$, one has 
$K_j=\frac{\partial}{\partial C_j}\equiv
\frac{\partial x^\mu}{\partial C_j}\frac{\partial}{\partial x^\mu}$. 
However,  strictly speaking $C_j$ are not coordinates on the 
group manifold ${\cal H}$ but rather on its tangent space 
at the group unity, such that their range is infinite. 
We wish to restrict the  range of $C_j$, and for this we should 
integrate not over the tangent space but over ${\cal H}$ itself.  
In other words, to render the integral in (\ref{42aa}) convergent 
we must treat the zero modes non-perturbatively, 
and for this we should replace the perturbative measure 
$\prod_j dC_j$ by a non-perturbative one, $d\mu(C)$. 

In general it is a difficult issue to construct the 
non-perturbative  path integration measure. 
However, in the 
zero mode sector this can be done. We note
that the measure should be invariant under the group 
multiplications, $d\mu(CC')=d\mu(C)$, and this uniquely
requires that  $d\mu(C)$ should be the Haar measure 
for ${\cal H}$. 
The normalization is fixed by the requirement
that for $C_j\to 0$ the perturbative result (\ref{42aa})
is reproduced. This unambiguously gives  
\be                                 \label{42aaa}
\Omega_1=\int
\left(\prod_j
\frac{\mu_o^2}{\sqrt{\pi}}\,
\left|\left|\frac{\partial}{\partial C_j}
\right|\right|\right)
d\mu(C)\, ,    
\ee   
where $\frac{\partial}{\partial C_j}$ is computed at $C_j=0$
and $d\mu(C)$ is the Haar measure of the isometry group ${\cal H}$
normalized such that 
$d\mu(C)\to \prod_j dC_j$ as $C_j\to 0$. 
 
\subsubsection{The path integral}

Now, using 
(\ref{30a}) and the measure (\ref{39}), 
we compute the path integral in (\ref{17}) -- first 
without the Faddeev-Popov factor ${\cal D}_{FP}$: 
\bea				
\int D[h_{\mu\nu}]
\exp\left(-\delta^2 I_{gf}\right)&=&
\Omega_2 \left(\prod_k^{\ \ \ \prime}
\frac{\mu_o}{\sqrt{\epsilon_k}}\right)\left(
\prod_s^{\ \ \ \prime}
\frac{\mu_o}{\sqrt{\gamma\sigma_s}} \right)\left(
\prod_n
\frac{\sqrt{2}\mu_o}
{\sqrt{\tilde{\lambda}^\gamma_n}}\right) \nonumber  \\
&\times&\frac{\mu_o}{\sqrt{2\Lambda}}\left(
\prod_i\frac{\sqrt{3\gamma}\,\mu_o}{\sqrt{2\Lambda}}
\right)\left(\prod_n
\frac{\sqrt{2\gamma}\mu_o}
{\sqrt{\tilde{\lambda}^\gamma_n}} \right) . 	\label{43}
\eea 
Here the primes indicate that zero and negative modes should
be omitted from the products. Zero vector modes 
do not contribute 
since they are not present in the 
path-integration measure (\ref{39}), and we assume that
there are no negative vector modes, since otherwise
the metric on the space of fluctuations would not be
positive definite. For tensor fluctuations negative 
and zero modes are present in the measure (\ref{39}),
and their overall contribution is collected in the factor
$\Omega_2$ in (\ref{43}). Let us further assume that 
there are no zero tensor modes, which is the case 
for the manifolds of interest. If negative
modes are also absent then $\Omega_2=1$. If
there is 
one negative tensor mode with  eigenvalue
$\varepsilon_{-}<0$, then 
\be					\label{44}
\Omega_2= \frac{\mu_o}{\sqrt{2\pi}}\int
\,dC_{-}^\phi \,\exp\left(-
\frac12\,\varepsilon_{-}\,(C_{-}^\phi)^{2}\right).
\ee
The integral is computed via the deformation
of the contour to the complex plane, which gives
the purely imaginary result
\be					\label{45}
\Omega_2=\frac{\mu_o}{2i\,\sqrt{|\varepsilon_{-}|}}\, ,
\ee
with the factor of 1/2 arising in the course of the 
analytic continuation \cite{Callan77}. 

Both the Faddeev-Popov factor in (\ref{41}) and the path
integral in (\ref{43}) depend on the gauge parameter 
$\gamma$. However, the $\gamma$-dependence
exactly cancels in their product, 
which provides a very good consistency check. 
In particular, the relative normalization of the integration measures
fixed by Eqs.(\ref{37}) and (\ref{38})
is important.
If we had 
divided each factor in the mode products in (\ref{36}) by $a\neq 1$,
then the resulting path integral  would be proportional to
$(\prod_n a)\sim a^{\zeta(0)}$ with $\zeta$ being the $\zeta$-function
of the $\gamma$-dependent operator $\tilde{\Delta}^\gamma_0$.
Thus, unless $a=1$, the result would be gauge-dependent.

We therefore finally obtain the
following expression for the 
path integral in (\ref{17}):
\be					\label{46}
Z[g_{\mu\nu}]=
\frac{\Omega_2}{\Omega_1}\,
\frac{\mu_o}{\sqrt{2\Lambda}}\left(
\prod_i\sqrt{\frac{2\Lambda}{3}}\frac{1}{\mu_o}
\right)\left(\prod_s^{\ \ \ \prime}
\frac{\sqrt{\sigma_s}}{\mu_o} \right)\left(
\prod_k^{\ \ \ \prime}
\frac{\mu_o}{\sqrt{\epsilon_k}}\right){\rm e}^{-I}
\ee
Here $\Omega_2$ is 
the contribution of the negative  
tensor mode, and $\Omega_1$ is the isometry factor. 
As we expected, the contribution of all unphysical scalar
modes has canceled from the result.  The only scalar modes
which do contribute are the several lowest lying modes which seem 
to be physically distinguished. 
These are the constant conformal mode giving rise to the factor
$\mu_o/\sqrt{2\Lambda}$, and the 5 `conformal
Killing scalars' which exist only in the $S^4$ case and give rise to the 
 product over $i$. 
The next two factors in (\ref{46}) is the contribution
of the transversal vector modes and the TT-tensor modes. 
Finally, $I=I[g_{\mu\nu}]$ is the classical action. 

In order to apply the above formula for $Z[g_{\mu\nu}]$ we need
the eigenvalues of the fluctuation
operators. Now we  shall determine the latter 
for the manifolds $S^2\times S^2$ and $S^4$.

\section{Spectra of fluctuation operators}
\setcounter{equation}{0}

In this section we  derive the spectra
of small fluctuations around the $S^2\times S^2$ and $S^4$ 
instantons. In the $S^2\times S^2$ case the problem is 
tackled via solving the differential equations. 
It turns out that in a suitable basis
the system of 10 coupled equations for the 
gravity fluctuations splits into
10 independent equations. 
The latter are solved in terms of 
spin-weighted spherical harmonics. In the $S^4$ case 
the equations do not decouple and 
the direct approach is not so transparent. However, the 
problem
can be conveniently analyzed with group theoretic methods,
which was done some time ago by Gibbons and Perry 
\cite{Gibbons78a}.  
We shall describe the group theory approach
in some detail -- with the same principal 
result as in \cite{Gibbons78a}.

\subsection{Fluctuations around the $S^2\times S^2$ instanton}

Let us consider the  metric of 
the $S^2\times S^2$ instanton background,
\be                                            \label{a2}
ds^2=\left.\left.\frac{1}{\Lambda}\right(
(d{\vartheta_1})^2+
\sin^2{\vartheta_1}\,(d{\varphi_1})^2+
(d{\vartheta_2})^2+
\sin^2\vartheta_2\,(d{\varphi_2})^2\right)\, .
\ee
It is convenient to set $\Lambda=1$ for the time being; 
at the end of calculations the $\Lambda$-dependence is 
restored 
by multiplying all eigenvalues with $\Lambda$. 
We introduce the complex tetrad
\bea                                     \label{a4}
{e}^1=\left.\left.\frac{1}{\sqrt{2}}\right(
d{\vartheta_1}+\frac{i}{\sin{\vartheta_1}}\,
d{\varphi_1}\,\right),\ \ \
{e}^2=\left.\left.\frac{1}{\sqrt{2}}\right(
d{\vartheta_1}-\frac{i}{\sin{\vartheta_1}}\,
d{\varphi_1}\,\right),
\nonumber \\
{e}^3=\left.\left.\frac{1}{\sqrt{2}}\right(
d{\vartheta_2}+\frac{i}{\sin{\vartheta_2}}\,
d{\varphi_2}\,\right),\ \ \
{e}^4=
\left.\left.\frac{1}{\sqrt{2}}\right(
d{\vartheta_2}-\frac{i}{\sin{\vartheta_2}}\,
d{\varphi_2}\,\right).
\eea
The metric in (\ref{a2}) splits as $g_{\mu\nu}=
e^{a}_{\ \mu}e^{b}_{\ \nu}\eta_{ab}$, where
the tetrad metric is
\be                                      \label{a5}
\eta^{ab}=g^{\mu\nu}e^{a}_{\ \mu}e^{b}_{\ \nu}
=
\left(\begin{array}{cccc}
0 & 1 & 0&0\\
1 & 0 & 0&0\\
0&0 & 0 & 1\\
0&0 & 1 & 0
\end{array}\right).
\ee

\subsubsection{Tensor modes}

First we consider the  eigenvalue problem 
\be                                            \label{a1}
-\nabla_\alpha\nabla^\alpha \phi_{\mu\nu}
-2R_{\mu\alpha\nu\beta}\,\phi^{\alpha\beta}=\varepsilon
\phi_{\mu\nu}\, ,
\ee
where
\be                                           \label{a1:0}
\nabla_\mu\phi^\mu_\nu=0,\ \ \ \phi^\mu_\mu=0\, .
\ee
We expand $\phi_{\mu\nu}$ with respect to the 
complex basis (\ref{a4}),
\be                                  \label{a6}
\phi_{\mu\nu}=e^{a}_{\ \mu}e^{b}_{\ 
\nu}\Phi_{ab}\, ,
\ee
insert this into (\ref{a1}) and project the 
resulting equation onto the basis (\ref{a4}) again.
Remarkably, the system of 10 coupled equations 
 splits then into 10 independent equations for
the 10 tetrad projections $\Phi_{ab}$. A partial 
explanation of this fact is the existence of two
different parity symmetries acting independently
on the two spheres.

It is convenient to introduce the operator
\be						\label{a7}
\hat{{\cal D}}[\s,\vartheta,\varphi]
=\frac{\partial^2}{\partial\vartheta^2}
+\cot\vartheta\,\frac{\partial}{\partial\vartheta}
+2i\s\,\frac{\cot\vartheta}{\sin\vartheta}+
\frac{1}{\sin^2\vartheta}\,
\frac{\partial^2}{\partial\phi^2}
-\s^2\cot\vartheta\ ,
\ee 
whose eigenfunctions are the
spin-weighted spherical harmonics 
$_{\s}Y_{jm}$ \cite{Goldberg67}, 
\be                                \label{a8}
\hat{{\cal D}}[\s,\vartheta,\varphi]\, 
_{\s}Y_{jm}(\vartheta,\phi)=
(\s^2-j(j+1))\ _{\s}\,Y_{jm}(\vartheta,\phi). 
\ee
Here 
$j$ and $m$ are such that  $j\geq |\s|$ and 
$-j\leq m \leq j$.
One has $_{\s}Y_{jm}=0$ for $j<|\s|$. 
[Notice that we use the bold-faced $\s$ for the 
spin weight.]
The following
relations between harmonics with different 
values of the spin weight $\s$ are useful: 
\be                                          \label{a9}
\hat{{\cal L}}^{\pm}[\s,\vartheta,\varphi]
\ _{\s}Y_{jm}=
\pm\sqrt{(j\pm \s)(j\mp \s+1)}\ 
_{\s\mp 1}Y_{jm}\, ,
\ee
where
\be                                     \label{a9:0}
\hat{{\cal L}}^{\pm}[\s,\vartheta,\varphi]=
 \frac{\partial}{\partial\vartheta}
\mp\frac{i}{\sin\vartheta}\,
\frac{\partial}{\partial\phi}
\pm \s\cot\vartheta\, .
\ee
The harmonics for a fixed $\s$ form
an orthonormal set on $S^2$. 

Using the above definitions, Eqs. (\ref{a1}) can be represented as
\be                    \label{a10}
\left(
\hat{\cal{D}}[\s^1_{ab},\vartheta_1,\varphi_1]+
\hat{\cal{D}}[\s^2_{ab},\vartheta_2,\varphi_2]
-(\s^1_{ab})^2-(\s^2_{ab})^2
+2+\varepsilon\right)
\,\Phi_{ab}=0 \, , 
\ee
where $1\leq a,b\leq 4$
(no summation over $a,b$). Here the nonzero elements of the 
symmetric matrices  $\s^1_{ab}$ and $\s^2_{ab}$ are 
\bea
&&\s^1_{11}=-\s^1_{22}=2,\ \ 
\s^1_{13}=\s^1_{14}=-\s^1_{23}=-\s^1_{24}=1\,,  \nonumber \\
&&\s^2_{33}=-\s^2_{44}=2,\ \
\s^2_{13}=\s^2_{23}=-\s^2_{14}=-\s^2_{24}=1\,.\ \ 
\eea
The solution to Eqs. (\ref{a10}) reads 
\be                 \label{a11}
\Phi_{ab}=C_{ab}\ 
_{\s^1_{ab}}Y_ {j_1m_1}({\vartheta_1},{\varphi_1})
\,_{\s^2_{ab}}Y_{j_2m_2}(\vartheta_2,\varphi_2),\ \ \  
\ee
with $C_{ab}$ being integration constants. 
The eigenvalue, $\varepsilon$, is the same for
all $\Phi_{ab}$: 
\be                                \label{a12}
\varepsilon=j_1(j_1+1)+j_2(j_2+1)-2\, .
\ee
This is essentially the sum of squares of the two SO(3)
angular momentum operators acting independently 
on the two spheres.  

Let us now count the degeneracy of the modes. 
For this one should take into account 
the additional conditions 
(\ref{a1:0}), which gives algebraic constraints
for the coefficients $C_{ab}$. We consider 
first the trace condition $\phi^\mu_\mu=0$. 
In the language of the tetrad projections 
this reduces to 
$\Phi_{12}+\Phi_{34}=0$, or equivalently to 
\be                                       \label{a13}
C_{12}+C_{34}=0\, .
\ee
Hence only 9 out of the 10 constants $C_{ab}$  
are independent. 

Next, we consider the Lorenz condition 
$\nabla_\sigma\phi^\sigma_\mu=0$. 
This implies  
\bea               \label{a14}
&&\hat{{\cal 
L}}^{-}[\s^1_{a1},\vartheta_1,\varphi_1]\,\Phi_{a1}+
\hat{{\cal L}}^{+}[\s^1_{a2},\vartheta_1,\varphi_1]\,\Phi_{a2}
\nonumber   \\
&+&
\hat{{\cal L}}^{-}[\s^2_{a3},\vartheta_2,\varphi_2]\,\Phi_{a3}+
\hat{{\cal 
L}}^{+}[\s^2_{a4},\vartheta_2,\varphi_2]\,\Phi_{a4}=0\, 
\eea
(no summation over $a$). 
Inserting the solution (\ref{a11}) and using the 
recurrence relations in (\ref{a9}), these
conditions reduce to algebraic constraints
\bea
{\kappa}_1\,C_{11}-{\alpha}_1 C_{12}+
\alpha_2\,(C_{13}-C_{14})=0\, , \nonumber \\
{\alpha}_1\,C_{12}-{\kappa}_1 C_{22}+
\alpha_2\,(C_{23}-C_{24})=0\, , \nonumber \\
{\alpha}_1\,(C_{13}-C_{23})+
\kappa_2\,C_{33}-\alpha_2\,C_{34}=0\, ,\nonumber \\
{\alpha}_1\,(C_{14}-C_{24})+
\alpha_2\,C_{34}-\kappa_2\,C_{44}=0\, .\label{a15}
\eea
Here $\alpha_\iota=\sqrt{j_\iota(j_\iota+1)}$ (with 
$\iota=1,2$) and 
$\kappa_\iota=\sqrt{(j_\iota+2)(j_\iota-1)}$ for $j_\iota\geq 
1$
while  $\kappa_\iota=0$ for $j_\iota=0$. 

For $j_\iota\geq 2$ 
(which corresponds to quadrupole
or higher deformations of each sphere) 
none of the coefficients $\alpha_\iota$, $\kappa_\iota$ 
vanish, and the algebraic constraints (\ref{a15}) 
reduce the number of independent coefficients $C_{ab}$ to 
five. 
This  gives  the 
degeneracy $d$:
\be                                 \label{a16}
j_1\geq 2,\ j_2\geq 2, \ \ \ \
d=5\,{(2j_1+1)}(2j_2+1)\, .
\ee 
The situation is different for small values of $j_\iota$. 
Consider, for example, the $j_1=j_2=0$ sector. 
Since the harmonics $_\s Y_{jm}$ vanish for $j<|\s|$, 
we must set in the solution (\ref{a11}) all  
$C_{ab}$'s to zero, apart from $C_{12}=-C_{34}$.
The Lorenz condition  (\ref{a14}) is then fulfilled. 
As a result, there is only one independent integration
constant, which yields 
\be                                 \label{a17}
{j_1}=j_2=0,\ \ \ \ d=1.
\ee 
Notice that in this case $\varepsilon=-2$.

In a similar way one can consider the sector where 
$j_1=0$ and $j_2=1$ (or $j_1=1$ and $j_2=0$), in which case
$\varepsilon=0$. One discovers then that the Lorenz 
constraints (\ref{a14}) require that all non-trivial 
coefficients
$C_{ab}$ mush vanish. As a result, 
\be                                 \label{a18}
j_1=0,\ j_2=1,\ {\rm or}\ \
j_2=1,\ j_1=0,\
\ \ \ \ d=0,
\ee 
which shows that there are no zero modes. 

Next, 
\be                                 \label{a19}
j_1=j_2=1,\ \ \ \ d={(2j_1+1)}(2j_2+1)=9;
\ee 
and finally
\bea                                 \label{a20}
&&{j}_1\geq 2,\ j_2=0,\ \ \ \ d={2j_1+1};\nonumber \\
&&{j}_1\geq 2,\ j_2=1,\ \ \ \ d=9{(2j_1+1)},
\eea 
which conditions are symmetric under 
interchanging $j_1$ and $j_2$. 

To recapitulate, the spectrum of the tensor fluctuations 
contains one negative mode and no zero modes.

\noindent
\subsubsection{Vector modes}
Let us now consider  the  eigenvalue problem 
\be                                            \label{a21}
(-\nabla_\alpha\nabla^\alpha -\Lambda)\eta_{\mu}=
\sigma\,\eta_{\mu}
\ee
subject to the condition
\be                                           \label{a22}
\nabla_\mu\eta^\mu=0
\ee
for the vector fluctuations on the $S^2\times S^2$ 
background. We again 
expand the fluctuations with
respect to the basis (\ref{a4}),
\be                                  \label{a23}
\eta_{\mu}=e^{a}_{\ \mu}\Psi_{a}, 
\ee
insert this into (\ref{a21}), and project back to 
the tetrad. Similarly as in the tensor case,
the equations  decouple to give 
\be                              \label{a24}
(
\hat{{\cal D}}[\s^1_a,\vartheta_1,\varphi_1]+
\hat{{\cal D}}[\s^2_a,\vartheta_2,\varphi_2]+
1+\sigma)
\,\Psi_{a}=0\, ,
\ee
where $1\leq a\leq 4$
(no summation over $a$), and nonzero coefficients read
$\s^1_1=-\s^1_2=\s^2_3=-\s^2_4=1$. 
The solution is 
\be                 \label{a25}
\Psi_{a}=C_{a}\, 
\,_{\s^1_{a}}Y_ {j_1m_1}({\vartheta_1},{\varphi_1})
\,_{\s^2_{a}}Y_{j_2m_2}(\vartheta_2,\varphi_2),\ \ \  
\ee
with $C_{a}$ being integration constants, 
and the eigenvalue is the same for all $\Psi_a$:
\be                                \label{a25:0}
\sigma={ j_1(j_1+1)}+j_2(j_2+1)-2\, .
\ee
The Lorenz condition, $\nabla_\sigma\eta^\sigma=0$, reads 
\bea               \label{a26}
&&\hat{{\cal L}}^{-}[\s^1_{1},
\vartheta_1,\varphi_1]\,\Psi_{1}+
\hat{{\cal L}}^{+}[\s^1_{2},\vartheta_1,\varphi_1]\,\Psi_{2}
\nonumber   \\
&+&
\hat{{\cal L}}^{-}[\s^2_{3},\vartheta_2,\varphi_2]\,\Psi_{3}+
\hat{{\cal 
L}}^{+}[\s^2_{4},\vartheta_2,\varphi_2]\,\Psi_{4}=0\, ,
\eea
which reduces upon inserting (\ref{a25}) 
to the algebraic condition
\be                                 \label{a27}
\sqrt{j_1(j_1+1)}\,(C_1-C_2)+
\sqrt{j_2(j_2+1)}\,(C_3-C_4)=0.
\ee 
This allows one to count the degeneracies:  
\be                                 \label{a28}
{j}_1\geq 1,\ j_2\geq 1\ \ 
(\sigma>0),\ \ d=3\,{(2j_1+1)}(2j_2+1);
\ee
and also 
\bea
&&j_1\geq 2 ,\ j_2=0\ \ (\sigma>0),
\ \ \ \ \  d=j_1(j_1+1); \nonumber \\
&&{j}_1=1,\ j_2=0\ \
(\sigma=0),\ \ \
\ \  d=3; \nonumber \\
&&{ j}_1=0,\ j_2=0\ \ (\sigma=-2),  \ \ \  d=0\, .
\eea 
These results are symmetric under $j_1\leftrightarrow j_2$,  
hence there are no negative modes, there are 
six zero modes corresponding
to the six Killing vectors of $S^2\times S^2$, 
and the rest of the spectrum is positive.

\subsubsection{Scalar modes and the orthogonality conditions}
The eigenvalue problem for the scalar modes,
\be                                            \label{a29}
-\nabla_\alpha\nabla^\alpha h=
\lambda\,h, 
\ee
reduces to the equation 
\be                              \label{a29a}
(
\hat{{\cal D}}[0,\vartheta_1,\varphi_1]+
\hat{{\cal D}}[0,\vartheta_2,\varphi_2]+
\lambda)
\,h=0\, ,
\ee
whose solutions are 
\be                 \label{a30}
h= 
\,Y_ {j_1m_1}({\vartheta_1},{\varphi_1})
\,Y_{j_2m_2}(\vartheta_2,\varphi_2) 
\ee
(for $\s=0$ the spin-weighted spherical harmonics 
coincide with the usual spherical harmonics). 
The eigenvalues are just  
\be                                \label{a31}
\lambda={j_1(j_1+1)}+j_2(j_2+1)\, ,
\ee
and the degeneracies are 
\be                                 \label{a32}
{j_1}\geq 0,\ j_2\geq 0,\ \ \ \ d={(2j_1+1)}(2j_2+1).
\ee

We have obtained the  spectra of all relevant fluctuation
operators. Although the eigenfunctions
are complex, one can  pick up their
real part in a way that is consistent with the 
orthogonality
conditions. For example, for the scalar modes one considers
\be                                     \label{a33}
\Re(h)=\frac{1+i}{\sqrt{2}}
\,\,{Y}_{j_1m_1}\,Y_{j_2m_2} + c.c \, ,
\ee
and one can easily see that the modes $\Re(h)$
with different quantum numbers $({j_1m_1}j_2m_2)$ are 
orthogonal with respect to the scalar product
defined in Eq. (\ref{11}). 

For the vector modes $\Psi_a$
the procedure is slightly more complicated, since the 
tetrad metric $\eta_{ab}$ is not diagonal. 
In addition, harmonics $_\s Y_{jm}$  
for different values of the spin weight are not 
orthogonal. Consider, however, the real combinations
\be                                  \label{a34}
\Re(\eta_{\mu})=\frac{1+i}{\sqrt{2}}\,
e^{a}_{\ \mu}\Psi_{a}+c.c\, ,  
\ee
where $\Psi_a$ has
quantum numbers $(j_1m_1j_2m_2)$. 
Consider 
$\Re(\eta^{(1)}_{\mu})$ and $\Re(\eta^{(2)}_{\mu})$
with different quantum numbers.
Their scalar product (defined in Eq. (\ref{11}))
can be computed using the relations 
\be                                  \label{a35}
\eta_{ab}=e_a^{\ \mu}e_b^{\ \nu}\,g_{\mu\nu},\ 
\ \ \
\delta_{ab}=e_a^{\ \mu}
(e_b^{\ \nu})^\ast\,g_{\mu\nu}\, ,\ \ 
\ee
which gives
\bea                                 \label{a36}
\langle \Re(\eta^{(1)}_{\mu}),\Re(\eta^{(2)\mu})\rangle=
\sum_a\langle\Psi^{(1)}_a,\Psi^{(2)\ast}_a\rangle \\
+
i\langle\Psi^{(1)}_1,\Psi^{(2)}_2\rangle+
i\langle\Psi^{(1)}_3,\Psi^{(2)}_4\rangle
+c.c.   \nonumber 
\eea
Here each term in the sum 
$\sum_a\langle\Psi^{(1)}_a,\Psi^{(2)\ast}_a\rangle$ 
is a bilinear combination
of spin-weighted  harmonics with 
{ the same} value of the spin weight, such that
the orthogonality relation holds.
Next, integrating by parts and using 
the recurrence relations (\ref{a9}) 
one can show that the remaining term in the scalar product, 
$i\langle\Psi^{(1)}_1,\Psi^{(2)}_2\rangle+
i\langle\Psi^{(1)}_3,\Psi^{(2)}_4\rangle+c.c$, 
vanishes. 
This shows that vector modes $\Re(\eta_{\mu})$ with
different quantum numbers are orthogonal.

\begin{table}
\caption{Spectra of fluctuations around the $S^2\times S^2$
instanton}
\vglue 0.4 cm
\begin{tabular}{|c|c|c|c|}\hline\hline
operator & eigenvalue & degeneracy &  \\ \hline
$\Delta_2$& $-2\Lambda$ & $1$ &  \\
& $2\Lambda$  & $9$ &   \\
& $(j(j+1)-2)\Lambda$ & $2(2j+1)$ & $j\geq 2$ \\
& $j(j+1)\Lambda$ & $18(2j+1)$ & $j\geq 2$ \\
& $(j_1(j_1+1)+j_2(j_2+1)-2)\Lambda$ &
$5(2j_1+1)(2j_2+1)$  &
$j_1,j_2\geq 2$  \\
\hline
$\Delta_1$ & 0 & 6 &  \\
& $(j(j+1)-2)\Lambda$ &
$2(2j+1)$  &
$j\geq 2$  \\
& $(j_1(j_1+1)+j_2(j_2+1)-2)\Lambda$ &
$3(2j_1+1)(2j_2+1)$  &
$j_1,j_2\geq 1$  \\
\hline
$\Delta_0$ & $(j_1(j_1+1)+j_2(j_2+1))\Lambda$ &
$(2j_1+1)(2j_2+1)$  &
$j_1,j_2\geq 0$  \\
\hline\hline
\end{tabular}
\end{table}

A similar
procedure can be carried out for the tensor modes.
Hence for all eigenmodes considered
above one can choose the real part in such a way that the 
orthogonality condition holds. This is 
a manifestation
of the fact that the fluctuation operators are 
self-adjoint. We finally restore the dependence on 
$\Lambda$ and summarize
the results of this section in Tab.1. There is one negative mode
in the spectrum, and this plays a crucial role in our
analysis. The corresponding deformation increases
the radius of one of the spheres, shrinking at the same time 
the second one.   

\subsection{Fluctuations around the $S^4$ instanton}

The $S^4$ instanton can be viewed as the 
four-dimensional 
sphere with radius $\sqrt{3/\Lambda}$ in 
five-dimensional Euclidean space $E^5$. 
Although 
the corresponding eigenvalue problem for fluctuations 
was considered in \cite{Gibbons78a}, 
we have reanalyzed it for the sake of completeness
(with the same result)
and shall present below the key steps of our analysis. 
The problem essentially reduces to studying 
representations 
of SO(5)
\cite{Hamermesh,Adler72,Adler73,Barut}, 
whose Casimir operator 
can be 
related to the invariant Laplacians on $S^4$ with the 
help 
of the projection formalism \cite{Hawking73}. We shall
therefore first outline the group theory part
by summarizing the relevant facts about representations
of $SO(5)$. 
We shall work on the unit 4-sphere rescaling at the 
end the eigenvalues by the factor $\Lambda/3$. 

\subsubsection{Representations of SO(5)}
The unit sphere $S^4$ in $E^5$
is defined in Cartesian coordinates 
by the equation $\sum_{a=1}^5(x^a)^2=1$.  
It is convenient to use  the complex coordinates 
$\xi^{\pm 1}=(x^1\pm i x^2)/\sqrt{2}$,
$\xi^{\pm 2}=(x^3\pm i x^4)/\sqrt{2}$, $\xi^0=x^5$.  
We shall not distinguish between lower and upper
indices, $\xi_i=\xi^i$.  
In these new coordinates the defining quadratic form 
reads  
$\sum_{i=-2}^2\xi^i\xi^{-i}=1$, which is annihilated
by the generators of SO(5): 
\be
Y^{i}_{j}=\xi^i\frac{\partial}{\partial \xi^j}- 
\xi^{-j}\frac{\partial}{\partial \xi^{-i}},\ \ \ \
\ee
whose commutation relations are 
\be
[Y^i_j,Y^k_l]=
\delta^{k}_j Y^i_l
-\delta^{i}_l Y^k_j
+\delta^{-l}_j Y^i_{-k}
-\delta^{k}_{-i} Y^{-j}_l\equiv C^{pq}_{ij,kl}Y^p_q\, .
\ee
Since $Y^{i}_{j}=-Y^{j}_{i}$, the independent
generators can be chosen to be those for $-i<j$. 
$Y^{1}_{1}$ and $Y^{2}_{2}$
generate the Cartan subalgebra, while $Y^{i}_{j}$
and $Y^{j}_{i}$ for $-i<j<i$ are the raising and 
lowering 
operators, respectively. 
One has
\be
[Y^i_i,Y^{k}_{l}]=\alpha^{k}_{l}(i)Y^{k}_{l}\, ,
\ee
where 
\be
\alpha^{k}_{l}(i)=
\delta^{i}_{k}- 
\delta^{i}_{l}+\delta^{i}_{-l}-\delta^{i}_{-k}
\ee 
determine the root vectors with the components 
$\alpha^{k}_{l}\equiv 
(\alpha^{k}_{l}(1),\alpha^{k}_{l}(2))$.
The roots corresponding to the 
four raising operators are
$\alpha^{2}_{1}=(-1,1)$, 
$\alpha^{2}_{0}=(0,1)$, 
$\alpha^{2}_{-1}=(1,1)$, and
$\alpha^{1}_{0}=(1,0)$. 

Irreducible representations of SO(5) are characterized 
by two numbers denoted by $m\equiv (m_1,m_2)$, where $m_2\geq m_1$
and both $m_1$ and $m_2$ 
are either integer or half-integer.  
The highest weight vector 
$\psi_m$ is annihilated by all raising operators, 
$Y^{i}_{j}\psi_m=0$ for $i>j>-i$, 
and it is an eigenvector of the Cartan subalgebra 
generators,
$Y^{i}_{i}\psi_m=m_i\psi_m$, $i=1,2$. 
Using these properties and also 
$[Y^i_j,Y^{j}_{i}]=Y^i_i-Y^j_j-2\delta^i_{-j}Y^i_{-j}$,
one finds the eigenvalues of the Casimir operator,
\be
\hat{C}\psi_m\equiv\frac12\sum_{i,j}Y^{i}_{j}Y^{j}_{i}\psi_
m
=C_m\psi_m\, ,
\ee
where 
\be                               \label{Casimir}
C_m=m_1(m_1+1)+m_2(m_2+3)
\ee
is the same for all vectors of the representation. 
The dimension of representations is given by 
\be
{\rm dim}(m)=\prod_\alpha\langle\alpha,r+m\rangle/
\prod_\alpha \langle\alpha,r\rangle\, . 
\ee
Here the product is over the four root vectors described
above, and $r=\frac12\sum_\alpha\alpha=(\frac12,\frac32)$.  
One has $r+m=(\frac12+m_1,\frac32+m_2)$, and 
$\langle\, ,\,\rangle$ is 
the scalar with respect to the Cartan metric 
 $g_{ij}=-C^{pq}_{ii,kl}C^{kl}_{jj,pq}=6\delta_{ij}$
(here $i,j=1,2$). As a result, 
\be                               \label{dim}
{\rm dim}(m)=\frac16\,(2m_1+1)(2m_2+3)
(m_2-m_1+1)(m_1+m_2+2)\, .
\ee

\subsubsection{Scalar modes}
Using Eqs.(\ref{Casimir}),(\ref{dim}) one can find the 
spectra
of the relevant fluctuation operators. 
It is now convenient to pass back to the Cartesian 
coordinates
$x^a=x_a$ ($a=1,\ldots 5$), 
such that the sphere $S^4$ is 
determined by the condition $r\equiv\sqrt{x^a x_a}=1$. 
Let $n^a\equiv x^a/r$ be the unit normal to the sphere. 
The (anti-hermitean) generators of SO(5) in 
Cartesian coordinates are given by 
$M_{ab}=n_a{\partial_b}- n_b{\partial_a}$, 
and the Casimir operator is 
$\hat{C}=-\frac12(M_{ab})^2\equiv
-\frac12 \sum_{ab}(M_{ab})^2$.

Let us define the projection operator 
$P_{ab}=\delta_{ab}-n_a n_b=P^{ab}=P^a_b$, 
which can be thought of as
the induced metric on the sphere. In what follows we 
shall 
use the projection formalism \cite{Hawking73} to
describe geometrical 
4-objects tangent to the sphere in terms of 
5-objects
of the embedding space. For example, a 4-vector 
$\eta_\mu$
can be described as a 5-vector $\eta_a$ subject to the 
condition $n^a\eta_a=0$. 
The covariant derivative of a tensor is  obtained
by taking the  partial derivative and then projecting
all the indices down to the sphere. For example, 
$\nabla_a\eta_b=(\partial_p\eta_q)P^p_a P^q_b$. 
One has $n_a=n^a$, while for objects tangent to the 
sphere 5-indices can be raised
and lowered either with $P_{ab}$ or with $\delta_{ab}$. 
The curvature tensor is given by 
$R_{psqt}=P_{pq}P_{st}-P_{pt}P_{sq}$. 

Consider first scalar fluctuations.
The covariant Laplacian 
for a scalar field $h$ can be expressed in terms of 
the 
angular momentum operator as
\be             \label{box:scal}
\Box h\equiv
P^{ab}\partial_a(P_b^c\partial_c h)=
\frac12 (M_{ab})^2 h=
-\hat{C}h\, .
\ee
Scalars transform according to the $(0,j)$ 
representations, 
which correspond to the Young tableaux 
\mbox{\begin{picture}(0,0)
\put (0,-1.5){\framebox(9,9){\tiny 1 }}
\put (9.5,-1.5){\framebox(9,9){\tiny 2 }}
\put (38,-1.5){\framebox(9,9){\tiny j}}
\put (0,-1.9) {\line(1,0){47}}
\put (0,7.6) {\line(1,0){47}}
\end{picture}
\mbox{\hspace{0.6cm}\ldots}\hspace{0.4cm}}
and can be represented in terms of homogeneous
polynomials on $S^4$ as 
\be
h=h_{(a_1\ldots a_j)}n^{a_1}\ldots n^{a_j}\,.     
\label{mode0}
\ee
Hence, the eigenvalues of the Casimir operator in 
Eq. (\ref{box:scal}) are
$C_{(0,j)}=j(j+3)$, which gives 
the spectrum of the scalar eigenvalue problem, 
$\Delta_0 h=\lambda h$ 
with $\Delta_0 \equiv -\frac{\Lambda}{3}\Box$: 
\be                  \label{spec0}
\lambda=\frac{\Lambda}{3}\,j(j+3),\ \ \ \
d=\frac16\,(2j+3)(j+2)(j+1),\ \ \ \  j\geq 0. 
\ee

\subsubsection{Vector modes}
Consider a tangent vector field $\eta_s=P^a_s\eta_a$.
The invariant Laplacian reads
\be                        \label{box:vec}
\Box\eta_s\equiv
P^{ab}\partial_a(P_b^c\partial_c \eta_p P^p_q)P^q_s=
\frac12 (M_{ab})^2
\eta_s+2(\partial_a\eta^a)n_s+\eta_s \, .
\ee
Here the last two terms 
on the right can be related to the contribution of the 
spin operator. 
Under general SO(5) rotations a vector $\eta(x)$
transforms into 
$\tilde{\eta}(x)=R\,\eta(R^{-1}x)$, where
$R=\exp(\omega^{ab}S_{ab})$ with 
$\omega^{ab}=-\omega^{ba}$
being the rotation parameters and $S_{ab}\equiv 
(S_{ab})^{pq}=
\delta_a^p\delta_b^q-\delta_b^p\delta_{a}^{q}$. 
For $|\omega_{ab}|\ll 1$ 
one obtains 
$\tilde{\eta}-\eta=\omega^{ab}(M_{ab}+S_{ab})\eta$, 
such that $S_{ab}$ can be identified with the spin 
operator:
$(S_{ab}\eta)_s=(S_{ab})_s^{~p}\eta_p$. 
As a result, 
\be
\Box\eta_s=\left(
\frac12 (M_{ab}+S_{ab})^2 +3\right)\eta_s\equiv
(-\hat{C}+3)\,\eta_s\, ,
\ee
where the Casimir operator is now the square of 
the total  angular momentum. 
The general vector harmonics on $S^4$ 
can be obtained by considering the product of a 
constant vector in $E^5$ with scalar
harmonics on $S^4$. Such a product decomposes into three
irreducible pieces, $(0,1)\otimes (0,j)=
(1,j)\oplus(0,j+1)\oplus(0,j-1)$,
which can be visualized as
\be               \label{young}
\mbox{\begin{picture}(0,0)
\put (0,-1.5){\framebox(9,9)}
\end{picture}}{\hspace{0.3cm}}
\otimes
\mbox{\begin{picture}(0,0)
\put (0,-1.5){\framebox(9,9){\tiny 1 }}
\put (9.5,-1.5){\framebox(9,9){\tiny 2 }}
\put (38,-1.5){\framebox(9,9){\tiny j}}
\put (0,-1.9) {\line(1,0){47}}
\put (0,7.6) {\line(1,0){47}}
\end{picture}
\mbox{\hspace{0.6cm}\ldots}\hspace{0.4cm}}
=
\mbox{\begin{picture}(0,0)
\put (0,-1.5){\framebox(9,9){\tiny 1 }}
\put (9.5,-1.5){\framebox(9,9){\tiny 2 }}
\put (38,-1.5){\framebox(9,9){\tiny j}}
\put (0,-10.9){\framebox(9,9)}
\put (0,-1.9) {\line(1,0){47}}
\put (0,7.6) {\line(1,0){47}}
\end{picture}
\mbox{\hspace{0.6cm}\ldots}\hspace{0.4cm}}
\oplus
\mbox{\begin{picture}(0,0)
\put (0,-1.5){\framebox(9,9){\tiny 0 }}
\put (9.5,-1.5){\framebox(9,9){\tiny 1 }}
\put (38,-1.5){\framebox(9,9){\tiny j}}
\put (0,-1.9) {\line(1,0){47}}
\put (0,7.6) {\line(1,0){47}}
\end{picture}
\mbox{\hspace{0.6cm}\ldots}\hspace{0.4cm}}
\oplus
\mbox{\begin{picture}(0,0)
\put (0,-1.5){\framebox(9,9){\tiny 1 }}
\put (9.5,-1.5){\framebox(9,9){\tiny 2 }}
\put (38,-1.5){\framebox(9,9){\tiny j-1}}
\put (0,-1.9) {\line(1,0){47}}
\put (0,7.6) {\line(1,0){47}}
\end{picture}
\mbox{\hspace{0.6cm}\ldots}\hspace{0.4cm}}\, .
\ee
The first term on the right here is the ($1,j$)-piece, 
and in the language 
of homogeneous polynomials 
it reads
\be                           \label{poly}
\eta_s=\eta_{[sa](a_1\ldots a_{j-1})}n^a n^{a_1}\ldots 
n^{a_{j-2}}
n^{a_{j-1}}\, ,
\ee
where $\eta_{[sa](a_1\ldots a_{j-1})}$
is traceless with respect to any pair of indices. 
This is manifestly tangential and coexact.
As a result, the eigenvalues of the Casimir operator 
are $C_{(1,j)}=j(j+3)+2$, and 
the spectrum of the vector eigenvalue problem 
$\Delta_1 \eta_s\equiv 
(-\frac{\Lambda}{3}\Box-\Lambda)\eta_s
=\sigma\eta_s$ in the sector where 
$\partial_a\eta^a=n^a\eta_a=0$ 
is given by 
\be                \label{spec1}
\sigma=\frac{\Lambda}{3}\,(j(j+3)-4),\ \ \ \
d=\frac12\,j(j+3)(2j+3),\ \ \ \ j\geq 1\, .
\ee
One can also directly verify that the harmonic $\eta_s$ in 
Eq.(\ref{poly}) fulfills the condition 
$\frac12(M_{ab})^2\eta_s=-j(j+3)\eta_s$.
It follows then from Eq. (\ref{box:vec}) that
$\Box\eta_s=-(j(j+3)-1)\eta_s$, and this again yields 
the spectrum 
in Eq. (\ref{spec1}). The 
correct degeneracy can be obtained
by counting the independent components 
of $\eta_{[sa](a_1\ldots a_{j-1})}$ 
\cite{Hamermesh}. 

The remaining two pieces in (\ref{young}), when 
represented in terms of the polynomials on $S^4$, 
can be related to the exact 
tangential and the normal components of the vector 
field.

\subsubsection{Tensor modes}
For a symmetric
tensor $h_{pq}=P^a_p P^b_q h_{ab}$ a direct calculation 
gives 
\bea                        \label{box:tens}
\Box h_{pq}+2R_{psqt}h^{st}&\equiv&
P^{ab}\partial_a(P_b^c(\partial_c h_{\bar{p}\bar{q}}) 
P^{\bar{p}}_{\underline{p}}P^{\bar{q}}_{\underline{q}})
P^{\underline{p}}_p P^{\underline{q}}_q 
+2(P_{pq}P_{st}-P_{pt}P_{sq})h^{st} \nonumber \\
&=&\frac12 (M_{ab})^2 h_{pq}
+2n_p(\partial^a h_{aq})+2n_q(\partial^a h_{ap})
+2\delta_{pq}h^a_a
\nonumber  \\
&=&
\left(
\frac12 (M_{ab}+\Sigma_{ab})^2 +6\right)h_{pq}\equiv
(-\hat{C}+6)h_{pq}\, .
\eea
Here the spin operator is defined in the same way as 
for vectors,
which gives
$(\Sigma_{ab}h)_{pq}=
(S_{ab})_{p}^{~s}h_{sq}+
(S_{ab})_{q}^{~s}h_{sp}$. 
The general tensor harmonics on $S^4$ are obtained by the 
direct products $(0,2)\otimes (0,j)=
(0,j+2)\oplus
(1,j+1)\oplus
(0,j)\oplus
(2,j)\oplus
(0,j+1)\oplus
(1,j-1)\oplus
(0,j-2)$.
Again this can be visualized by Young's diagrams and 
represented
in the language of homogeneous polynomials. 
The transverse and tracefree  harmonics 
tangent to the sphere correspond to
the $(2,j)$ piece, whose explicit representation is 
\be
h_{pq}=h_{[pa][qb](a_1\ldots a_{j-2})}n^a n^b 
n^{a_1}\ldots n^{a_{j-2}}\, .         \label{mode2}
\ee
Here $h_{[pa][qb](a_1\ldots a_{j-2})}$ is traceless
with respect to any pair of indices and is symmetric
under interchange of the $[pa]$ and $[qb]$ pairs.  
As a result, the 
eigenvalues of the Casimir operator are 
$C_{(2,j)}=j(j+3)+6$. 
This gives 
the spectrum of the tensor eigenvalue problem 
$\Delta_2 h_{pq}=\varepsilon h_{pq}$ in the sector where
$\partial^ah_{ab}=n^ah_{ab}=h^a_a=0$:  
\be                         \label{spec2}
\varepsilon=\frac{\Lambda}{3}\,j(j+3),\ \ \ \
d=\frac56\,(j-1)(j+4)(2j+3),\ \ \ \ j\geq 2\, .
\ee
The same result can be obtained by directly verifying 
that $h_{pq}$
in Eq. (\ref{mode2}) fulfills the condition
$\frac12(M_{ab})^2 h_{pq}=-j(j+3) h_{pq}$.

The other tensor harmonics in the expansion of 
$(0,2)\otimes (0,j)$
correspond to the exact and coexact pieces of the 
longitudinal vector part of the 4-metric, to those of the 
4-vector $h_{5\mu}$, and to the trace \cite{Gibbons78a}.

We summarize the results of our analysis in this section in Tab.2. 
Notice that the scalar and tensor eigenvalues are the 
same
(for $j\geq 2$), while the vector spectrum is shifted
by a constant. 

\begin{table}
\begin{center}
\caption{Spectra of fluctuations around the $S^4$
instanton}
\vglue 0.4 cm
\begin{tabular}{|c|c|c|c|}\hline\hline
operator & eigenvalue & degeneracy &  \\ \hline
$\Delta_2$& $\frac{\Lambda}{3}j(j+3)$ & 
$\frac56(j-1)(j+4)(2j+3)$ & $j\geq 2$ \\
\hline
$\Delta_1$& $\frac{\Lambda}{3}(j(j+3)-4)$ & 
$\frac12 j(j+3)(2j+3)$ & $j\geq 1$ \\
\hline
$\Delta_0$& $\frac{\Lambda}{3}j(j+3)$ & 
$\frac16(j+1)(j+2)(2j+3)$ & $j\geq 0$ \\
\hline\hline
\end{tabular}
\end{center}
\end{table}

\section{Partition function}
\setcounter{equation}{0}
Now we are able to derive the explicit
expressions for the one-loop partition functions
for fluctuations around the $S^2\times S^2$ and 
$S^4$ instantons. The corresponding formula
was obtained in Eq.(\ref{46}) above. It is 
convenient to pass from $\mu_o$ to the dimensionless 
normalization parameter $\muo$ via the rescaling 
\be                                      \label{a39}
\mu_o=\sqrt{\Lambda}\muo\, . 
\ee
The one-loop
partition function for gravity fluctuations around
an Euclidean background  then reads
\be					\label{a40}
Z[g_{\mu\nu}]=
\frac{\muo}{\sqrt{2}}\,\Omega_0\,  
\frac{\Omega_2}{\Omega_1}\,
\sqrt{\frac{{\rm Det}^\prime\Delta_1}
{{\rm Det}^\prime\Delta_2}}\, 
{\rm e}^{-I}\, .
\ee
Here the first two factors on the right
are the contributions of the scalar modes. 
The factor  $\muo/\sqrt{2}$  
is due to the constant conformal mode, 
which is always present, and    
$\Omega_0$ is the contribution of the
5 scalar modes with eigenvalue
$4\Lambda/3$ which exist only for the $S^4$
instanton (see Tab.2):    
\be
\Omega_0=
\left(\sqrt{\frac{2}{3}}\frac{1}{\muo}\right)^5\, .
\ee
For any background other than $S^4$ one has $\Omega_0=1$.  
As was discussed above, other scalar modes do not
contribute to the partition function.

The factor $\Omega_2$ in (\ref{a40})
is the contribution of the negative tensor mode,   
\be					\label{a41}
\Omega_2=\frac{\muo}{2i\,\sqrt{|\varepsilon_{-}|}}\, .
\ee
For the $S^2\times S^2$ instanton there is one 
such mode 
with $\varepsilon=-2$, 
while in 
the $S^4$ case all tensor modes are positive and 
$\Omega_2=1$. Next, 
\be                                 \label{a41a}
\Omega_1=
\left(\prod_j
\frac{\mu_0^2}{\sqrt{\pi}}\,\Lambda\,
\left|\left|\frac{\partial}{\partial C_j}
\right|\right|\right)
Vol({\cal H})\, ,    
\ee   
is the isometry factor.  
If the background has no isometries then $\Omega_1=1$.

The determinants in Eq. (\ref{a40})
are the contributions of the positive 
vector and tensor modes. One has 
\be                    \label{a42}
\sqrt{{\rm Det}^\prime\Delta_1}=
\left(\prod_s^{\ \ \ \prime}
\sqrt{\frac{{\sigma_s}}{{\Lambda}\muo^2}} 
\right)
=\exp\left(-\frac12\zeta_1^\prime(0)
-\frac12(\ln\muo^2)\,\zeta_1(0)\right)\, ,    
\ee
where the $\zeta$-function for the positive, transverse
vector modes is 
\be                   \label{a43}
\zeta_1(z)=\sum_s\!^\prime
\left(\frac{\Lambda}{\sigma_s}\right)^z\, .
\ee
Similarly for the positive,  transverse traceless
tensor modes: 
\be                     \label{a44}
\sqrt{{\rm Det}^\prime\Delta_2}=
\left(\prod_k^{\ \ \ \prime}
\sqrt{\frac{{\epsilon_k}}{{\Lambda}\muo^2}} 
\right)
=\exp\left(-\frac12\zeta_2^\prime(0)
-\frac12(\ln\muo^2)\,\zeta_2(0)\right)   
\ee
with 
\be                     \label{a45}
\zeta_2(z)=\sum_k\!^\prime
\left(\frac{\Lambda}{\epsilon_k}\right)^z\, .
\ee
The last factor in Eq. (\ref{a40}) is the 
classical contribution, with $I$ being the action 
for the background. Let us now apply
these formulas. 

\subsection{The $S^2\times S^2$ instanton}
The classical action is
$I[S^2\times S^2]=-2\pi/\Lambda G$,
and according to Tab.1, 
\be					\label{a46}
\Omega_0=1\,,\ \ \ \ \ 
\Omega_2=\frac{\muo}{2i\,\sqrt{2}}\, .
\ee
Consider now the isometry factor $\Omega_1$ in (\ref{a41a}), 
which is due to the background SO(3)$\times$SO(3) symmetry.   
Each of the two SO(3) groups can be parameterized by matrices
$U_{ik}=\exp(\varepsilon_{ikj}C_j$).    
The invariant metric on the SO(3) space is 
$g_{ik}=\frac12 {\rm tr}(\partial_i U\partial_k U^{-1})\to \delta_{ik}$
for $C_j\to 0$. The Haar measure is
$d\mu(C)=\sqrt{{\rm det} g_{ik}}\,dC_1 dC_2 dC_3$, and the volume
$Vol$(SO(3))= $\int d\mu(C)=8\pi^2$. For later use,  we reproduce
this result in a different way.  
The measure for a (compact, semi-simple) Lie group ${\cal G}$
can be represented as 
the product of the measure for the maximal subgroup ${\cal H}$ 
and that for the coset ${\cal G}/{\cal H}$. This implies that 
\be                         \label{coset}
Vol({\cal G})=Vol({\cal H})\times Vol({\cal G}/{\cal H})\, .
\ee
In particular, $Vol$(SO(3))=$Vol$(SO(2))$\times Vol(S^2)$, 
where $Vol$(SO(2))= $2\pi$, and the volume of the 
$S^2$ coset with unit (due to the normalization of the 
measure) radius is $Vol(S^2)=4\pi$. 
As a result, $Vol$(SO(3))=$2\pi\times 4\pi=8\pi^2$. 

When acting on $S^2$, the SO(3) generators 
$\frac{\partial}{\partial C_j}$ 
generate rotations  in the three orthogonal planes of the 
embedding Euclidean 3-space.  
Let $\frac{\partial}{\partial C_3}$ be the
generator of rotations in the XY-plane, such that the azimuthal
angle of the spherical coordinate system 
changes as $\varphi\to\varphi+C_3$. 
Then the norm
$||\frac{\partial}{\partial C_3}||$  
is the square root of 
\be                       \label{a46a}
\langle\frac{\partial}{\partial\varphi},
\frac{\partial}{\partial\varphi}\rangle=
\frac{1}{32\pi G}\int_{S^2\times S^2}g_{\varphi\varphi}
\sqrt{g}d^4x=\frac{\pi }{3\Lambda^3 G}\, .  
\ee
Obviously, the norms $||\frac{\partial}{\partial C_1}||$ 
and $||\frac{\partial}{\partial C_2}||$ 
and those of the generators of the second SO(3) factor 
are the same. Hence,   
\be                                  \label{a46b}
\Omega_1=
\left(\frac{\mu_0^2}{\sqrt{\pi}}\,\Lambda 
\left|\left|\frac{\partial}{\partial\varphi}\right|\right|
\right)^6 \left(Vol({\rm SO(3)})\right)^2
=\frac{64\pi^4(\mu_0)^{12}}{27(\Lambda G)^3}\, .
\ee

Consider now the positive modes.
The $\zeta$-function associated with 
the positive vector modes is (see Tab.1)
\be                            \label{a47}
\zeta_1(s)=
\sum_{j=2}^\infty
\frac{2(2j+1)}{\{j(j+1)-2\}^s}
+
\sum_{j_1=2}^\infty\sum_{j_2=2}^\infty
\frac{3(2j_1+1)(2j_2+1)}{\{j_1(j_1+1)+j_2(j_2+1)-2\}^s}.
\ee
This can be represented as
\be                            \label{a48}
\zeta_1(s)=4^s\,(2\,\zeta(2,-9|s)+3\,Z(1,-10|s)),
\ee
where the following two functions have been introduced:
\bea                       \label{a49}
\zeta(k,\nu|s)&=&\sum_{j=k}^\infty
\frac{2j+1}{\{(2j+1)^2+\nu\}^s}\,, \\
Z(k,\nu|s)&=&\sum_{j_1=k}^\infty 
\sum_{j_2=k}^\infty\,\frac{(2j_1+1)\,(2j_2+1)}
{\left\{(2j_1+1)^2+(2j_2+1)^2+\nu\right\}^s}\, . \label{a49aa}
\eea 
These functions are studied in detail in the Appendix. 
Similarly, using the results of Tab.1 one obtains 
the $\zeta$-function for the positive tensor
modes
\be                           \label{a50}
\zeta_2(s)=9\times 2^{-s}+
4^s\,(2\,\zeta(2,-9|s)+18\,\zeta(2,-1|s)
+5\,Z(2,-10|s)). 
\ee
The following relation  implied by the definitions in (\ref{a49}), 
(\ref{a49aa}), will be useful:
$Z(1,-10|s)=Z(2,-10|s)+6\zeta(2,-1|s)+9\times 8^{-s}$. 

\subsubsection{The scaling behaviour}

Before we proceed further, 
it is very instructive to pause and check
whether the expressions above agree with the general 
formulas for the  scaling behaviour of effective actions. 
We shall follow the approach of Christensen and
Duff \cite{Christensen80}, who relate this scaling behaviour to
\bea
N_0&=&\frac{1}{180\,(4\pi)^2 }\int (
R_{\mu\nu\rho\sigma}R^{\mu\nu\rho\sigma}+636\Lambda^2)
\sqrt{g}\,d^4x\, , \nonumber \\
N_1&=&\frac{1}{180\,(4\pi)^2}\int (
-11R_{\mu\nu\rho\sigma}R^{\mu\nu\rho\sigma}+984\Lambda^2)
\sqrt{g}\,d^4x\, , \nonumber \\
N_2&=&\frac{1}{180\,(4\pi)^2}\int (
189 R_{\mu\nu\rho\sigma}R^{\mu\nu\rho\sigma}-756\Lambda^2)
\sqrt{g}\,d^4x\, . \label{a51}
\eea
Here $N_0$ is the `number of eigenvalues' of the 
scalar operator $\Delta_0-2\Lambda$ acting on a 
manifold with $R_{\mu\nu}=\Lambda g_{\mu\nu}$. 
$N_1$ is the number of eigenvalues of the vector
operator $\Delta_1$ acting in the space of all 
vectors, that is, including both transverse and
longitudinal fluctuations. Finally, $N_2$ counts
both transverse and longitudinal
eigenstates of the tensor operator $\Delta_2$, 
with the only requirement that the fluctuations
must be traceless. 

Let us apply these formulas to the $S^2\times S^2$
background. The volume of the manifold is 
$V_{S^2\times S^2}=(4\pi)^2/\Lambda^2$, while 
$R_{\mu\nu\rho\sigma}R^{\mu\nu\rho\sigma}=8\Lambda^2$. 
As a result,
\be                  \label{a52}
N_0=\frac{161}{45}\, ,\ \ \ 
N_1=\frac{224}{45}\, ,\ \ \ 
N_2=\frac{21}{5}\, .
\ee
Now let us obtain the same result via a direct
evaluation of the $\zeta$-functions. 
First we consider the scalar case. 
Using the results of Tab.1, 
the  operator $\Delta_0-2\Lambda$ has 
one negative 
mode, six zero modes, while the rest 
of the spectrum is positive and 
gives rise to the $\zeta$-function
\be                         \label{a53}
\zeta_0(s)=4^s\,(2\,\zeta(2,-9|s)+\,Z(1,-10|s))\, .
\ee
Hence the number of all eigenvalues is
$7+\zeta_0(0)$. In order to compute $\zeta_0(0)$,
we use the results of the Appendix, where 
the following values are obtained: 
\bea                       \label{a54}
\zeta(k,\nu|0)&=&\frac{1}{12}-\frac14\,\nu-k^2\,,\ \ \\
Z(k,\nu|0)&=&\frac{1}{32}\,\nu^2-\frac{1}{24}\,\nu+2k^4
+(\frac12\,\nu-\frac23)\,k^2+\frac{13}{360}\,.
\eea
This gives for the $\zeta$-functions in (\ref{a48}),
(\ref{a50}), (\ref{a53}) 
\be                       \label{a55}
\zeta_0(0)=-\frac{154}{45}\,,\ \ \  
\zeta_1(0)=-\frac{18}{5}\, ,\ \ \ \
\zeta_2(0)=\frac{38}{9}.
\ee 
Using these, the number of scalar
eigenvalues is $N_0=7-\frac{154}{45}=\frac{161}{45}$,
which agrees with (\ref{a52}). 

Next, the vector operator $\Delta_1$ has 6 zero modes, 
such that the number of its eigenvalues in the transverse
sector is $6+\zeta_1(0)$. Now, one should take into
account also the longitudinal vectors, 
which are gradients of scalars.   
It is not difficult to see that if $\nabla_\mu\chi$
is an eigenvector of $\Delta_1$, such that 
$\Delta_1\nabla_\mu\chi=\sigma\nabla_\mu\chi$, then 
$(\Delta_0-2\Lambda)\chi=\sigma\chi$. 
We see that the eigenfunctions of
$\Delta_0-2\Lambda$ are in  one-to-one 
correspondence with the longitudinal vectors.  
The number of the latter is therefore $N_0-1$,
where the one is subtracted because the ground state
scalar eigenfunction is constant, which vanishes
upon differentiation. We therefore conclude that
$N_1=6+\zeta_1(0)+N_0-1=6-\frac{18}{5}+\frac{161}{45}-1=
\frac{224}{45}$, which also agrees with (\ref{a52}). 

Finally, the number of traceless eigenvalues of
$\Delta_2$ is $1+\zeta_2(0)$ (here the one is the contribution
of the  negative mode) plus the number of longitudinal
traceless tensor harmonics 
$\phi^{\rm L}_{\mu\nu}=\nabla_\mu\xi_\nu+  
\nabla_\nu\xi_\mu-\frac12\,g_{\mu\nu}\nabla_\rho\xi^\rho$.
 
Now, if $\Delta_1\xi_\mu=\sigma\xi_\nu$
then for $\phi^{\rm L}_{\mu\nu}$ associated with $\xi_\mu$
one has 
$\Delta_2\phi^{\rm L}_{\mu\nu}=\sigma\phi^{\rm 
L}_{\mu\nu}$. 
Hence, the number of longitudinal tensors is 
determined by the number of vectors, which gives
$N_2=1+\zeta_2(0)+(N_1-6)$. Here six is subtracted
because the six Killing vectors do not contribute 
to the tensor spectrum, since for Killing vectors one has
$\phi^{\rm L}_{\mu\nu}=0$. We therefore obtain 
$N_2=1+\frac{38}{9}+\frac{224}{45}-6=\frac{21}{5}$, which 
again is in perfect agreement with (\ref{a52}).

The overall scale dependence
of the partition function 
is given by the factor $(\mu_0)^{N_2+N_0-2N_1}$.  
For the $S^2\times S^2$ instanton one has  
$N_2+N_0-2N_1=-\frac{98}{45}$, and we shall shortly see 
that this agrees with our analysis. 

\subsubsection{The partition function $Z[S^2\times S^2]$}

It is now a simple task to insert
the formulas above into the expression for the partition
function. We obtain
\be                         \label{a56}
\sqrt{\frac{{\rm Det}^\prime\Delta_1}
{{\rm Det}^\prime\Delta_2}}\
=\exp\left(\zeta'(0)
+\ln\muo^2\,\zeta(0)\right)\, ,
\ee
where
\be
\zeta(s)\equiv\frac12\,(\zeta_2(s)-\zeta_1(s))=
-9\times 2^{-s}+4^s\,Z(2,-10|s)\, .
\ee
Using the values $Z(2,-10|0)=\frac{581}{45}$ and 
$Z'(2,-10|0)\equiv\Gamm=-18.3118$ (see Eq.(\ref{ZETA})
in the Appendix) we find
\be                         \label{a56a}
\sqrt{\frac{{\rm Det}^\prime\Delta_1}
{{\rm Det}^\prime\Delta_2}}\
=2^{\frac{1567}{45}}\muo^{\frac{352}{45}}{\rm 
e}^\Gamm\, .
\ee
Finally, taking into account the contributions of the
negative, zero, and scalar modes computed in 
(\ref{a46}),  
together with the classical term, we obtain 
\bea                     \label{a57}
Z[S^2\times S^2]=
-i\,\frac{27\,(\Lambda G)^3}{256\,\pi^4\mu_{0}^{10}}
\sqrt{\frac{{\rm Det}^\prime\Delta_1}
{{\rm Det}^\prime\Delta_2}}\, {\rm e}^I \\
=
-i\, 0.3667\times (\Lambda G)^3
\muo^{-\frac{98}{45}}
\exp\left(\frac{2\pi}{\Lambda G}\right).  \nonumber
\eea
This is our final result in the $S^2\times S^2$ sector. 

\subsection{The $S^4$ instanton}
The classical action is
$I[S^4]=-3\pi/\Lambda G$.
Using the results in Tab.2 we find
\be					\label{a58}
\Omega_0=
\left(\sqrt{\frac{2}{3}}\frac{1}{\muo}\right)^5\, ,
\ \ \ \ \ \
\Omega_2=1\, .
\ee
Let us consider the symmetry factor $\Omega_1$. 
The isometry group is now ${\cal H}$=SO(5), and this
can be represented by matrices $U_{ik}=\exp(C_{ik})$,
where $C_{ik}=-C_{ki}$, $i,k=1,\ldots 5$. The 10 generators 
$\frac{\partial}{\partial C_{ik}}$ generate rotations of $S^4$ 
in the 10 orthogonal planes of the 
embedding Euclidean 5-space. 
Let $\frac{\partial}{\partial C_{12}}$ be the generator of 
rotations in the XY-plane, such that the standard azimuthal angle
changes as $\varphi\to\varphi+C_{12}$. The norm 
$||\frac{\partial}{\partial C_{12}}||$ is the square root of 
\be                       \label{a46aa}
\langle\frac{\partial}{\partial\varphi},
\frac{\partial}{\partial\varphi}\rangle=
\frac{1}{32\pi G}\int_{S^4}g_{\varphi\varphi}
\sqrt{g}\,d^4x=\frac{9\pi }{10\Lambda^3 G}\, ,
\ee
which applies also to the the norms of the remaining 9 generators.

The volume of SO(5) can be computed by directly constructing the 
invariant metric and the Haar measure with the use of the matrix 
representation $U_{ik}=\exp(C_{ik})$. 
The measure should be normalized such that for 
$C_{ik}\to 0$ it reduces to $\prod_{i<k}dC_{ik}$. 
However, it is much simpler to use the coset reduction formula (\ref{coset}). 
One has SO(5)/SO(4)=$S^4$ and SO(4)/SO(3)=$S^3$, 
such that $Vol$(SO(5))=$Vol(S^4)\times Vol(S^3)\times Vol$(SO(3)). 
We know that $Vol$(SO(3))=$8\pi^2$, while the volumes of unit $S^3$ 
and $S^4$ are $2\pi^2$ and $8\pi^2/3$, respectively.  
As a result, $Vol$(SO(5))=$128\pi^6/3$. Summarizing,  
\be                                  \label{a46ba}
\Omega_1=
\left(\frac{\mu_0^2}{\sqrt{\pi}}\,\Lambda 
\left|\left|\frac{\partial}{\partial\varphi}\right|\right|
\right)^{10} Vol({\rm SO(5)})
=\left(\frac{9}{10}\right)^5\,\frac{128\pi^6}{3}\,
\frac{(\mu_0)^{20}}{(\Lambda G)^5}\, .
\ee

Let us consider the positive modes. 
The $\zeta$-function associated with 
the positive vector modes is (see Tab.2)
\be                            \label{a59}
\zeta_1(s)=\frac12\,3^s
\sum_{j=2}^\infty
\frac{j(j+3)(2j+3)}{\{j(j+3)-4\}^s}\, .
\ee
This can be written as
\be                            \label{a60}
\zeta_1(s)=\frac12\,3^s\,
{\cal Q}(1,-4,0|s)\, ,
\ee
where the following function has been introduced
\be                       \label{a61}
{\cal Q}(k,\nu,c|s)=\sum_{j=k}^\infty
\frac{(2j+3)(j(j+3)+c)}{\{j(j+3)+\nu\}^s}\,, 
\ee
Similarly, using the results of Tab.2, one obtains 
the $\zeta$-function for the positive tensor
modes
\be                           \label{a62}
\zeta_2(s)=\frac56\,3^s\,{\cal Q}(2,0,-4|s)\, .
\ee
Finally, consider the scalar operator 
$\Delta_0-2\Lambda$. According to Tab.2, its
eigenvalues, measured in units of $\Lambda$,
are given by $(j(j+3)-6)/3$, and the degeneracy is
$(j+1)(j+2)(2j+3)/6$ with $j\geq 0$. Hence,
the $\zeta$-function for the positive scalar
modes is
\be                           \label{a63}
\zeta_0(s)=\frac16\,3^s\,{\cal Q}(2,-6,-4|s)\, .
\ee

\subsubsection{The scaling behaviour}

Let us again check the consistency with the general 
formulas for the  scaling behaviour of quantum fields 
(for fluctuations around $S^4$ this was done by Christensen and
Duff \cite{Christensen80}).  
Applying again the formulas in (\ref{a51}), where
now the volume of the manifold is 
$V_{S^4}=24\pi^2/\Lambda^2$, while 
$R_{\mu\nu\rho\sigma}R^{\mu\nu\rho\sigma}=8\Lambda^2/3$,  
one has
\be                  \label{a64}
N_0=\frac{479}{90}\, ,\ \ \ 
N_1=\frac{358}{45}\, ,\ \ \ 
N_2=-\frac{21}{10}\, .
\ee
On the other hand, using the result of
the Appendix, 
\bea                       \label{a65}
{\cal Q}
(k,\nu,c|0)&=&-\frac12\,k^4-2k^3-(c+\frac12)k^2 \\
&+&(3-2c)k+\frac32\,\nu^2+\frac13\,c
-\frac{11}{15}\, , \nonumber 
\eea
one obtains for the $\zeta$-functions in (\ref{a59}),
(\ref{a60}), (\ref{a62}) 
\be                       \label{a66}
\zeta_0(0)=-\frac{61}{90}\,,\ \ \  
\zeta_1(0)=-\frac{191}{30}\, ,\ \ \ \
\zeta_2(0)=-\frac{61}{90}.
\ee 
Now, since the spectrum of $\Delta_0-2\Lambda$ contains
six non-positive modes, one has $N_0=6+\zeta_0(0)=
6-\frac{61}{90}=\frac{479}{90}$, which agrees with 
(\ref{a64}). Next, $\Delta_1$ has 10 zero modes, 
such that there are $10+\zeta_1(0)$ transverse vector
eigenstates, plus $(N_0-1)$ longitudinal ones 
(the constant scalar mode gives no contribution).
As a result, $N_1=10-\frac{191}{30}+\frac{479}{90}-1=
\frac{358}{45}$, which agrees with (\ref{a64}). 
Finally, there are $N_2=\zeta_2(0)+N_1-15$ traceless
tensor modes, where 15 is subtracted because 10
Killing vectors and 5 conformal Killing vectors
of $S^4$ do not contribute to the longitudinal tensor
modes. One obtains 
$N_2=-\frac{61}{90}+\frac{358}{45}-15=-\frac{21}{10}$,
which again agrees with (\ref{a64}). 

The overall scale dependence of the partition functions 
is expected to be $(\mu_0)^{N_2+N_0-2N_1}$, where
$N_2+N_0-2N_1=-\frac{571}{45}$.

\subsubsection{The partition function $Z[S^4]$}

Let us now obtain the partition
function. One finds  
\be                         \label{a67}
\sqrt{\frac{{\rm Det}^\prime\Delta_1}
{{\rm Det}^\prime\Delta_2}}\
=\exp\left(\zeta'(0)
+\ln\muo\,\zeta(0)\right)\, ,
\ee
where
\be                       \label{a68}
\zeta(s)\equiv\frac12\,(\zeta_2(s)-\zeta_1(s))=
3^s\left(\frac{5}{12}\,{\cal Q}(2,0,-4|s)-
\frac{1}{4}\,{\cal Q}(2,-4,0|s)\right).
\ee
One has $\zeta(0)=\frac{509}{90}$ and  
$\zeta'(0)\equiv\Gamm_1=6.1015$ (see Eq.(\ref{b21g}) in the Appendix). 
This yields 
\be                         \label{a56aa}
\sqrt{\frac{{\rm Det}^\prime\Delta_1}
{{\rm Det}^\prime\Delta_2}}\
=\muo^{\frac{509}{45}}{\rm e}^{\Gamm_1}\, .
\ee
Finally, collecting the contributions of the
negative, zero, and scalar modes computed in 
(\ref{a46}),  
together with the classical term, we obtain 
\bea                     
Z[S^4]=
\frac{\sqrt{3}\,5^5}{3^{12}\pi^6\mu_0^{24}}\,
\sqrt{\frac{{\rm Det}^\prime\Delta_1}
{{\rm Det}^\prime\Delta_2}}\, {\rm e}^I  \nonumber \\
=
0.0047\times (\Lambda G)^5
\muo^{-\frac{571}{45}}
\exp\left(\frac{3\pi}{\Lambda G}\right).  \label{a57a}
\eea
To our knowledge, this formula has been obtained here for the first time,
since in Refs.\cite{Gibbons78a,Christensen80}
 a closed expression for $Z[S^4]$ was not achieved. In particular,
the isometry factor $\Omega_1$ was not taken into account
and the derivative of the $\zeta$-function was not computed.

\section{Summary}
\setcounter{equation}{0}

Our last step is to use the expressions for $Z[S^2\times S^2]$
and $Z[S^4]$ in (\ref{a57}) and (\ref{a57a}) 
and insert these into Eq.(\ref{rate}) to find the decay rate
\bea                     
\Gamma&=&-\frac{1}{\pi}\,\sqrt{\frac{\Lambda}{3}}\,
\frac{\Im Z[S^2\times S^2]}{Z[S^4]} 
\nonumber \\
&=& 
14.338\,\sqrt{\Lambda}\, 
(G\Lambda)^{-2} (\mu_o\Lambda)^{\frac{473}{45}}
\exp\left(-\frac{\pi}{\Lambda G}\right).         \label{final}
\eea
This is the final result of our analysis. This formula gives 
the rate of semiclassical decay of de Sitter space due to the
spontaneous nucleation of black holes. This is the leading
mode of decay, since classically de Sitter
space is stable \cite{Ginsparg83}. 
The numerical coefficient in the formula originates from 
the  fluctuation
determinants evaluated in the $\zeta$-function scheme. 
The factor $\sqrt{\Lambda}$ comes from the heat bath
temperature coefficient in (\ref{rate}) and gives $\Gamma$
the correct dimension of an inverse time. The coefficient 
$(G\Lambda)^{-2}$ arises due to the combined effect of the
background isometries. 
The power of $\mu_o\Lambda$ contains the effect of rescalings,
where we have passed again to the dimensionful renormalization
parameter $\mu_o$. Since quantum gravity is 
non-renormalizable, $\mu_o$ remains undetermined, 
and we have nothing to say about this problem. 
For numerical estimates it is reasonable to assume 
that $\mu_o\sim G$. The last factor in the formula is the 
classical term. The formula is obtained
in the one-loop approximation, which is good as long as the 
classical term is large compared to the quantum corrections, that is 
for $\Lambda G\ll 1$. Under this condition the nucleation
rate is exponentially small. Notice that since the overall power
of $\Lambda$ is positive, the quantum corrections provide 
an additional suppression of the transition rate for small $\Lambda$.

The formula gives the probability of black hole nucleation 
per unit proper time of a freely falling 
observer in his Hubble region. The latter is the region 
enclosed inside the observer's cosmological horizon.
If a black hole is created, then it has the radius $1/\sqrt{\Lambda}$ 
and fills the whole Hubble region. This does not mean that 
the whole space will be eaten by a giant black hole, since
de Sitter spacetime consists of many Hubble regions, whose number
grows as the universe expands. Some of these regions will
contain a black hole but most of them will be empty. 
The black holes are actually born in pairs, where the two 
members of the pair are created at the opposite sides 
of the 3-space.  The interesting conclusion is that for $G\Lambda\ll 1$, when 
inflation is `slow', the rate of black hole nucleation is strongly
suppressed, but the created black holes are large. This can be 
understood as a consequence of the fact that the black holes 
are made of the energy contained inside the Hubble region. As the 
size of the latter is large for small $\Lambda$, the created black 
holes are also large. On the other hand, if one is allowed to 
extrapolate the formula for  $G\Lambda\sim 1$, when  
inflation is fast, then the created black holes are small, but they
are created in abundance.   

One can see that for late times the number of 
black holes per unit physical volume will be constant. 
Let us choose for de Sitter spacetime the global coordinates associated
with the freely falling observers:
\be                           \label{dSglobal0}
ds^2=-d\eta^2+\frac{3}{\Lambda}\cosh^2\left(
\sqrt{\frac{\Lambda}{3}}\,\eta\right)\,d\Omega_3^2\, .
\ee
Here $\eta$ is the (dimensionful) proper time and $d\Omega_3^2$
is the volume element of the unit 3-sphere. The volume of the 
global hypersurface $\Sigma_\eta$ of constant $\eta$ is 
$V(\eta)=
2\pi^2\left(\frac{3}{\Lambda}\right)^{3/2}
\cosh^3\left(\sqrt{\frac{\Lambda}{3}}\,\eta\right)\approx 
\frac{\pi^2}{4}\,\left(\frac{3}{\Lambda}\right)^{3/2}
\exp(\sqrt{3\Lambda}\eta)$. The portion of $\Sigma_\eta$
contained inside the future event horizon of any observer
has the volume $V_{\rm H}=
\frac{4\pi}{3}\left(\frac{3}{\Lambda}\right)^{3/2}$ (for late 
$\eta$). This is the spatial Hubble volume.
[This quantity slightly depends on the choice of the 
hypersurface. Even though for any given observer one has 
$\eta=t$, which is the time associated 
with the observer's coordinate system, one has
$\Sigma_\eta\neq\Sigma_t$, unless 
$\eta=t=0$, in which case the spatial Hubble volume is
$V_{\rm H}=\pi^2\left(\frac{3}{\Lambda}\right)^{3/2}$].  
As a result, the number of Hubble volumes on the hypersurface
is $N_{\rm H}(\eta)=V(\eta)/V_{\rm H}$. 
[One has $N_{\rm H}(0)=2$: the de Sitter throat consists of 
two causally disconnected parts belonging to the Hubble regions of 
two antipodal observers \cite{Schrodinger56}.]
Multiplying $N_{\rm H}(\eta)$ by $\Gamma$ gives the black hole 
nucleation rate per $\Sigma_\eta$\,,
\be
\frac{dN_{\rm BH}}{d\eta}=\frac{3\pi}{16}
\exp(\sqrt{3\Lambda}\eta)\,\Gamma\, .
\ee
Integrating with respect to $\eta$ and dividing 
by $V(\eta)$ yields the average 
volume density of created black holes
on $\Sigma_\eta$,
\be
\rho_{\rm BH}=\frac{\Lambda}{12\pi}\, \Gamma\, ,
\ee
which does not depend on $\eta$. 

The subsequent real time evolution
of these black holes is an interesting issue. Presumably most of them will
immediately evaporate, unless $\Lambda$ is very small and the black holes
are large, in which case however the nucleation rate is strongly suppressed. 
It was argued in \cite{Bousso98} that this process could dramatically 
change the global structure of de Sitter space. 
For more information on this issue we refer
to \cite{Bousso96,Bousso98,Elizalde99}
and to the papers cited in Ref.\cite{Bousso98}. 

The following steps have been essential in our analysis.  
We have derived Eq.(\ref{rate}) for the 
nucleation rate using the thermal properties of  de Sitter space.
For this 
we have approximated  the partition function for Euclidean quantum 
gravity
with $\Lambda>0$ by the semiclassical contributions of the $S^4$
and $S^2\times S^2$ instantons, 
of which the first yields the free energy $F$ in 
the Hubble volume while the contribution of the 
second can be regarded as a purely imaginary part of $F$.  
In a sense one can think of the created  black holes as being the
bubbles of the new phase spontaneously created out of 
thermal fluctuations via quantum tunneling. 
We have argued that these bubbles may have 
temperature different from that of the heat bath, since they 
cannot thermalize via interactions with the whole reservoir
and only exchange energy 
inside the Hubble region. 

To compute the one-loop contributions of the $S^4$ and 
$S^2\times S^2$ instantons we have used the standard Faddeev-Popov
approach to the path integral. 
We have worked with a one-parameter family of covariant background gauges
and employed the Hodge decomposition of the fluctuations with their 
subsequent spectral expansion. In our treatment of the conformal
modes we have followed the standard recipe of  complex rotation,
up to several lowest lying modes for which a
different prescription has been applied.  
In order to integrate over zero modes of the Faddeev-Popov
operator arising due to the background isometries, we have gone beyond
the perturbation theory and showed that the corresponding 
integration measure is the Haar measure on the isometry group. 
There are no other zero modes in the problem -- for example, the standard
rotational zero modes are absent because rotations are isometries
of the backgrounds under consideration. 

We have explicitly determined the spectra of the fluctuation operators. 
For fluctuations around the $S^2\times S^2$ instanton the spectrum
was obtained by directly solving the differential equations, while 
in the $S^4$ case  group theoretic methods have been applied,
in which we followed the approach of \cite{Gibbons78a}. 
These spectra have been used in order to 
compute the functional determinants within the 
$\zeta$-function regularization scheme, the 
corresponding $\zeta$-functions being studied in detail
in the Appendix below. We have checked that our results agree
with the general formulas for the anomalous scaling behaviour. 
Finally, we have obtained in  (\ref{a57}), (\ref{a57a}) 
the one-loop partition functions for fluctuations around 
the $S^4$ and $S^2\times S^2$ backgrounds. 
To our knowledge, in both cases such closed expressions
have been obtained for the first time. 
The last step has been to use the resulting partition functions in 
order to calculate the nucleation rate $\Gamma$.
This describes a constant density of created black holes
per unit physical volume of the expanding 3-space. 

After the  work of Gross, Perry and Jaffe \cite{Gross82},
our analysis presents the second example of a 
complete one-loop computation on a non-trivial background.%
\footnote{Note also that the analysis in \cite{Gross82} was not quite 
complete, since the spectrum  is unknown and  the 
$\zeta$-functions have not been computed, even though the undetermined
quantities can be absorbed into the renormalization parameter.
 We also do not understand their treatment of the 
background isometries and that of the non-normalizable
deformations of the instanton.}
One may hope that our results can lend 
further  support to the Euclidean approach to quantum gravity.

\subsection*{Acknowledgments}
We thank Michael Bordag for suggesting the idea
to use the Abel-Plan formula in the analytic continuation
of the $\zeta$-functions. M.S.V. would also like to thank
Gary Gibbons for discussing the role of the special 
conformal modes and Raphael Bousso for interesting
conversations. The work of M.S.V. was supported by the 
Deutsche Forschungsgemeinschaft, grant Wi 777/4-2. 

{\bf Note added in proof.}
We would like to thank Dima Vassilevich for bringing
to our attention a number of relatively recent papers 
considering one-loop Euclidean quantum gravity on $S^4$. 
Although in none of these papers a closed expression for the 
one-loop partition function $Z[S^4]$ is achieved, 
it is worth mentioning the work by Allen \cite{Allen86}, 
by Polchinski \cite{Polchinski89}, and 
by Taylor and Veneziano \cite{Taylor90}. 
We refer to the paper by Vassilevich \cite{Vassilevich93} 
for more references. 
Not all  papers agree on the scaling behaviour 
of the partition function. The reason is that  
some authors do not take into 
account the contribution of the 10 zero modes 
due to the background isometries, 
thereby obtaining $Z[S^4]$ to be proportional to 
$\mu_0^{+\frac{329}{45}}$ instead of  
$\mu_0^{-\frac{571}{45}}$ \cite{Taylor90}. 
However, since these zero modes are in the path integration measure, 
they do contribute to the anomalous scaling on equal footing with 
all other modes. In fact, the example of  
flat space gauge theories \cite{INSTANTONS} 
shows that the background symmetry 
zero modes, when treated non-perturbatively as was done above, 
are of vital importance for obtaining the
correct running behaviour of the coupling constant.   
Our result for the scaling
behaviour agrees with that of 
Christensen and Duff \cite{Christensen80} 
and with the general analysis
of Fradkin and Tseytlin \cite{Fradkin84}.

\section*{Appendix. Calculation of $\zeta$-functions.}
\renewcommand{\theequation}{A.\arabic{equation}}
\renewcommand{\thesubsection}{A.\arabic{subsection}}
\setcounter{equation}{0}
In this Appendix we shall study the $\zeta$-function
\be                                   \label{A1}
Z(k,\nu|s)=\sum_{n=k}^\infty 
\sum_{m=k}^\infty\,\frac{(2n+1)\,(2m+1)}
{\left\{(2n+1)^2+(2m+1)^2+\nu\right\}^s}\, ,
\ee 
which is used in the main text for computing 
the one-loop fluctuation term on 
the $S^2\times S^2$ instanton background. 
Here $\nu$ is real while $k$ is a positive integer 
such that $2(2k+1)^2+\nu>0$. It is assumed that $\Re(s)$
is positive and large enough to ensure the convergence of 
the series. 
Despite its apparent simplicity, 
the analysis of this $\zeta$-function is lacking in the 
literature.
This is probably due to the fact that 
the summation in (\ref{A1})
cannot be extended to all integers
and the standard Poisson resummation
techniques do not apply. For this reason we 
use other methods, which are unfortunately 
rather lengthy. However we think that it is   
necessary to describe the basic steps, especially in view of
other possible applications of our results. 

In what follows we shall perform the analytic continuation
by finding the integral representation for $Z(k,\nu|s)$ that is
valid for any $s$. This will be 
used to compute the values of 
$Z(k,\nu|0)$ and 
$\frac{d}{ds}Z(k,\nu|s)$ at $s=0$. 
As a first step, we shall 
consider the related $\zeta$-function:
\be                                   \label{A2}
{\zeta}(k,\nu|s)=\sum_{n=k}^\infty 
\frac{(2n+1)}
{\left\{(2n+1)^2+\nu\right\}^s}\, 
\ee 
with $(2k+1)^2+\nu>0$. 
The integral representation for this function will be 
useful. 
In addition, we shall study
the $\zeta$-function 
\be                       \label{A61}
{\cal Q}(k,\nu,c|s)=\sum_{j=k}^\infty
\frac{(2j+3)(j(j+3)+c)}{\{j(j+3)+\nu\}^s}\,, 
\ee
where $k(k+3)+\nu>0$, and shall find its value and  
its $s$-derivative at $s=0$. This function 
is needed in the analysis of fluctuations
around the $S^4$ instanton.

\subsection{Computation of $Z(k,\nu|0)$ and 
${\zeta}(k,\nu|0)$.} 
First we shall compute the values of 
these functions at $s=0$ using the standard heat 
kernel technique. 
These values determine the scaling properties of the 
system. 
Later we shall rederive the same values by using
the integral representations for $Z(k,\nu|s)$
and ${\zeta}(k,\nu|s)$, 
and this will provide us with a good consistency check. 
For ${\cal Q}(k,\nu,c|s)$ we shall consider only the 
integral representation, since 
the values of ${\cal Q}(k,\nu,c|0)$ have been computed in 
\cite{Christensen80}.   

A $\zeta$-function  
related to a second order elliptic operator with a positive spectrum
can be expressed as
\be                                        \label{A3}
\zeta(s)=\frac{1}{\Gamma(s)}\int_0^\infty t^{s-1}
\Theta(t)\, dt\, .
\ee
On compact spaces the heat kernel 
$\Theta(t)$ vanishes exponentially fast for large $t$,
while for small $t$ there is the asymptotic expansion
\be                                             \label{A4}
\Theta(t)\sim\sum_r\, C_r\, t^r\, ,
\ee
with $r$ assuming in general both integer and half-integer
values. It is not difficult to see that
\be						\label{A5}
\zeta(0)=C_0\, .
\ee
The problem therefore reduces to determining the 
asymptotic expansion of the heat kernel.
The heat kernels in our problem are given by 
\be						\label{A6}
\Theta(k,\nu|t)=\left(\theta(t)-\xi(k|t)\left)^2\,
{\rm e}^{-\nu t}\right.\right.
\ee
for $Z(k,\nu|s)$ and 
\be						\label{A7}
{\theta}(k,\nu|t)=\left(\theta(t)-\xi(k|t)\left)\,
{\rm e}^{-\nu t}\right.\right.
\ee
for ${\zeta}(k,\nu|s)$, where
\be						\label{A8}
\theta(t)=\sum_{n=0}^\infty\, (2n+1)\,{\rm 
e}^{-t\,(2n+1)^2}\, 
\ee
and 
\be						\label{A9}
\xi(k|t)=\sum_{n=0}^{k-1}\, (2n+1)\,
{\rm e}^{-t\,(2n+1)^2}\, . 
\ee
The only difficulty is to find the asymptotic
expansions for small $t$ for the function $\theta(t)$ in 
(\ref{A8})%
\footnote{We note that $\theta(t)$ cannot be expressed
in terms of theta-functions in a simple way, and that
the Poisson resummation formula does not
directly apply.}. $\theta(t)$ is a partition
function for a two-dimensional rotator at temperature 
$1/t$. 
We wish therefore to find its high-temperature
expansion, and for this we shall  construct
the integral representation for $\theta(t)$.

Let us consider the ``generating function'' 
\be						
\label{A10}
\chi(t,\alpha)=\sum_{n=0}^\infty
{\rm e}^{-t\,(2n+1)^2+i\alpha\,(2n+1)}\, 
\ee
such that
\be						
\label{A11}
\theta(t)=
-i\lim_{\alpha\to 0}\frac{\partial}
{\partial\alpha}\,\chi(t,\alpha)\, .
\ee
$\chi(t,\alpha)$ fulfills the differential 
equation 
\be						
\label{A12}
\frac{\partial\chi}{\partial t}=
\frac{\partial^2\chi}{\partial\alpha^2}\, .
\ee
This has the special solution 
\be					     \label{A13} 
\tilde{\chi}(t,\alpha)=\frac{1}{\sqrt{4\pi t}}\, 
\exp\left(-\frac{(\alpha-\alpha_0)^2}{4t}\right)
\ee
with the property
$\tilde{\chi}(0,\alpha)=\delta(\alpha-\alpha_0)$, 
which allows us to represent the general solution of 
(\ref{A12}) as
\be					 \label{A14}
\chi(t,\alpha)=\int_{-\infty}^{\infty}
\tilde{\chi}(t,\alpha_0)\,
{\chi}(0,\alpha_0)\,d\alpha_0\, .
\ee
The initial value ${\chi}(0,\alpha_0)$ 
is obtained directly from the definition (\ref{A10}):
\be						
\label{A15}
\chi(0,\alpha_0)=\sum_{n=0}^\infty
{\rm e}^{i\alpha_0\,(2n+1)}\, =\frac{i}{2\sin\alpha_0}\, ,
\ee
where we assume that $\alpha_0$
has a small positive imaginary part in order 
to ensure convergence of the geometrical series. 
We can now insert this into (\ref{A14}) and the result 
into (\ref{A11}). Introducing the new variable 
$x=\alpha_{0}^2/4$ we  obtain 
the sought for integral representation 
\be						
\label{A16}
\theta(t)=\frac{1}{\sqrt{4\pi t^3}}\,
\int_0^\infty{\rm 
e}^{-x/t}\,\frac{dx}{\sin(2\sqrt{x})}\, .
\ee
Here we should remember that $x$ has a small 
imaginary part, such that the integration is actually
performed along a contour parallel to
the positive real axis and approaching it from above. 

It is now a straightforward task to find 
the asymptotic expansion of the integral in (\ref{A16}) 
for 
small $t$,
since the only non-trivial contribution comes from 
a small  neighbourhood  of $x=0$: 
\be                                        \label{A18}
\theta(t)\sim \frac{1}{4t}\,
\left(1+\frac13\, t+\frac{7}{30}\,t^2+O(t^3)\right).
\ee
Inserting this into (\ref{A6}) and (\ref{A7}) gives 
the asymptotic expansions for the heat kernels
$\Theta(k,\nu|t)$ and  ${\theta}(k,\nu|t)$,
whose coefficients $C_0$ determine the  
$\zeta$-functions at $s=0$:
\bea						
\label{A19}
Z(k,\nu|0)=
\frac{1}{32}\,\nu^2-\frac{1}{24}\,\nu+\frac{1}{2}\, k^2\nu
+2k^4-\frac23\,k^2+\frac{13}{360}\, ,
\eea
and 
\bea						
\label{A20}
{\zeta}(k,\nu|0)=\frac{1}{12}-\frac14\,\nu-k^2\, .
\eea
To check these results we note that 
the definitions in (\ref{A1}) and (\ref{A2}) imply that 
\bea					\label{A21}
Z(k_1,\nu|s)&=&Z(k_2,\nu|s)+
2\sum_{m=k_1}^{k_2-1}(2m+1){\zeta}(k_2,\nu+(2m+1)^2|s)
\nonumber \\
&+&\sum_{n=k_1}^{k_2-1} 
\sum_{m=k_1}^{k_2-1}\,\frac{(2n+1)\,(2m+1)}
{\left\{(2n+1)^2+(2m+1)^2+\nu\right\}^s}\, ,
\eea
with $k_2>k_1$. Setting here $s=0$ we obtain a 
non-trivial 
relation for $Z(k,\nu|0)$ and ${\zeta}(k,\nu|0)$, 
and this is fulfilled by the expressions in 
(\ref{A19}) and (\ref{A20}). 
Finally we use (\ref{A19}), (\ref{A20}) to obtain
the values used in the main text:
\bea
Z(2,-10|0)=\frac{581}{45}\, ,\ \ \  
{\zeta}(2,-9|0)=-\frac{5}{3}\, ,\ \ \ 
{\zeta}(2,-1|0)=-\frac{11}{3}\,  .   \label{A22}
\eea

\subsection{Computation of  ${\zeta}(k,\nu|s)$, and 
${\cal Q}(k,\nu,c|s)$.
} 

It is usually more difficult to determine   
the derivative of a $\zeta$-function at $s=0$
than the value of the function itself, 
since the knowledge of its behaviour in a
neighbourhood of $s=0$ is required.  
We shall perform the analytic continuation of the 
$\zeta$-functions defined by Eqs. (\ref{A1}), (\ref{A2}) 
and 
(\ref{A61}) to arbitrary values of $s$ 
with the use of the relation sometimes called

\subsubsection{The Abel-Plan formula.}
\begin{figure}[th]
\begin{minipage}[t]{13cm}
{\epsfysize=6 cm\epsffile{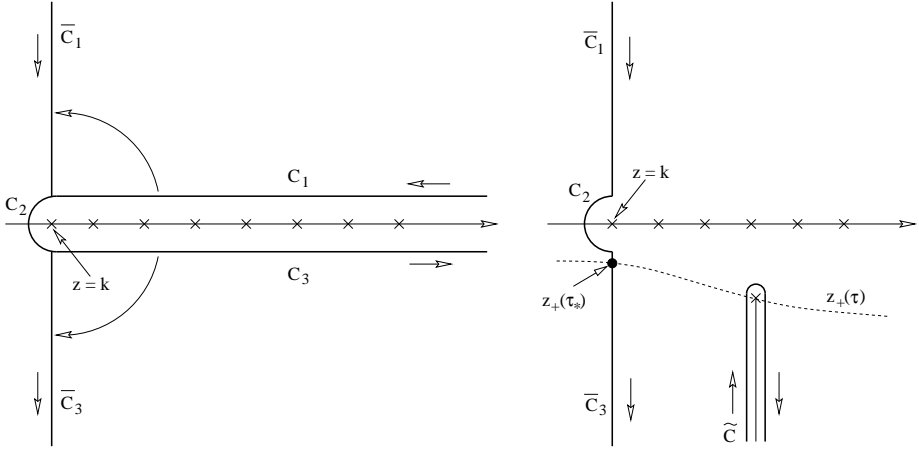}}
\caption
{
\label{FIG-CONTOUR}
{{\sl 
{\rm Left}: 
Starting from the contour $C_1+C_2+C_3$
and rotating 
we arrive at $\bar{C}_1+C_2+
\bar{C}_3$. {\rm Right}: the same when a 
branching point at $z=z_{+}(\tau)$ 
is present. The contour $C_1$ will then wrap
around the cut leading to the additional contribution due
to $\tilde{C}$. The point $z=z_{+}(\tau)$ is in the 
region of interest for $\tau\geq \tau_\ast$.}
}}
\end{minipage}
\end{figure}

This can be derived using  the obvious relation 
\be				\label{b1}
\sum_{n=k}^\infty f(n)=\int_C
\, \frac{f(z)}{{\rm e}^{2\pi iz}-1}\, dz\, ,
\ee
where the contour $C$
encompasses the part of the real axis with
$Re(z)\geq k$ (see Fig.\ref{FIG-CONTOUR})
 and $f(z)$ 
is analytic for $Re(z)\geq k$. 
The idea is to split $C$ into three parts,
$C_1+C_2+C_3$, as shown in 
Fig.\ref{FIG-CONTOUR}. For the first part, $C_1$, the 
integral can be written as 
\be				\label{b2}
\int_{C_1}
\, \left(\frac{1}{1-{\rm e}^{-2\pi iz}}-1\right)f(z)\, dz=
\int_k^\infty f(t)\, dt +
\int_{C_1}
\,\frac{f(z)}{1-{\rm e}^{-2\pi iz}}\, dz\, ,
\ee
where in the integral over $C_1$ on the right the contour 
is then rotated to the position $\bar{C}_1$ 
as shown in Fig.\ref{FIG-CONTOUR}. Such a rotation 
is possible if only $f(z)$ tends to zero fast enough
for $Re(z)\geq k$ and $|z|\to\infty$. 

The integral over the second portion of the contour,
$C_2$, is equal to $\frac12 f(k)$, while in the 
integral over $C_3$ the contour is rotated 
to the position $\bar{C}_3$ 
as shown in Fig.\ref{FIG-CONTOUR}. As a result, we arrive at
the Abel-Plan formula
\be				\label{b3}
\sum_{n=k}^\infty f(n)=\frac12\, f(k)
+\int_k^\infty f(t) dt
+i\int_0^\infty\frac{f(k+it)-f(k-it)}
{{\rm e}^{2\pi t}-1}\, dt\, .
\ee
This formula can be  used for analytic
continuation of $\zeta$-functions, in which case 
$f(t)$ depends also on $s$, $f=f(t,s)$. The 
analytic continuation to small values of 
$s$ is performed in the
first integral on the right in (\ref{b3}). This usually
converges only for $\Re(s)$ large and positive, but 
can often be computed in a closed form, and then one can
continue the result to arbitrary $s$.  
The second integral   
on the right in (\ref{b3}) usually cannot be computed
in a closed form, but it converges for
any $s$. Let us first apply the Abel-Plan formula 
to the $\zeta$-functions in (\ref{A2}) and 
(\ref{A61}). 

\subsubsection{Analytic continuation of $\zeta(k,\nu|s)$.}
Applying (\ref{b3}) to the series for $\zeta(k,\nu|s)$
in (\ref{A2}), we have  
\be					\label{b4}
f(z)=\frac{2z+1}
{\left\{(2z+1)^2+\nu\right\}^s}\, ,
\ee 
which 
is analytic for $\Re(z)>-1/2$ and decays fast enough
for $|z|\to\infty$ provided that $\Re(s)$ is large enough.
As a result, we can use the Abel-Plan formula, 
which gives 
\bea					\label{b5}
\zeta(k,\nu|s)=
\left(k+\frac12\right)
\frac{1}{\{(2k+1)^2+\nu\}^{s}}+
\frac{1}{4(s-1)}\frac{1}{\{(2k+1)^2+\nu\}^{s-1}} \nonumber 
\\
+\int_0^\infty\frac{idt}
{{\rm e}^{2\pi t}-1}
\left(
\frac{2k+1+2it}{\left\{(2k+1+2it)^2+\nu\right\}^s}-
\frac{2k+1-2it}{\left\{(2k+1-2it)^2+\nu\right\}^s}
\right).\ \ 
\eea
This representation is finite  
for all $s$, apart from $s=1$, where the pole is located.   
The remaining integral  
here converges uniformly for $|s|<\infty$,
which allows us to differentiate
with respect to $s$. If we set 
$s=0$, then the integral  can be easily computed. 
We find
${\zeta}(k,\nu|0)=\frac{1}{12}-\frac14\,\nu-k^2$, 
and this agrees with the value obtained above in 
(\ref{A20}).

Similarly,  
we can differentiate (\ref{b5}) with respect to 
$s$ and then set $s=0$. This gives
\bea					\label{b9}
\zeta'(k,\nu|0)=\frac14\,W(\ln W-1)
-\left(k+\frac12\right)\ln W        \nonumber \\
+2\int_0^\infty\frac{dt}
{{\rm e}^{2\pi t}-1}\,
(t\,\ln{\cal A}+(2k+1)\Psi)\, ,
\eea
where $W=(2k+1)^2+\nu$ and 
\be					\label{b8}
{\cal A}=(W-4t^2)^2+16(2k+1)^2 t^2\, ,\ \ \ \ 
\Psi=\arctan\frac{4(2k+1)\,t}{W-4t^2}\, .
\ee
For any $k$ and $\nu$ 
the integral in (\ref{b9}) is convergent and can be  
evaluated
numerically. Notice that 
$\zeta'(k,\nu|0)$ is not needed in the main body of the paper,
and for this reason we do not quote the actual number here. 

\subsubsection{Analytic continuation of 
$Q(k,\nu,c|s)$.}

The procedure is exactly the same as above.
Denoting 
\be                                 \label{b12a}
f(z)=\frac{(2z+3)(z(z+3)+c)}{\{z(z+3)+\nu\}^s}
\ee
the direct application of the Abel-Plan formula (\ref{b3}) 
yields
\bea
Q(k,\nu,c|s)=\left(k+\frac32\right)(k(k+3)+c)\,W^{-s}+
\frac{1}{s-2}\, W^{2-s} \nonumber \\
+\frac{c-\nu}{s-1}\, W^{1-s}+
\int_0^\infty\frac{dt}
{{\rm e}^{2\pi t}-1}\, i(f(k+it)-f(k-it))\, ,
\eea
where $W=k(k+3)+\nu$. Setting  $s=0$ the integral
can be easily computed leading to 
\bea                       \label{b21b}
{\cal Q}
(k,\nu,c|0)&=&-\frac12\,k^4-2k^3-\left(c+\frac12\right)k^2 \\
&+&(3-2c)\,k+\frac12\,\nu^2+\left(\frac43-\nu\right)c
-\frac{11}{15}\, .   \nonumber 
\eea
Next, differentiating (\ref{b12a}) with respect to $s$ and
setting $s=0$ gives 
\bea                          
&&Q'(k,\nu,c|0)=-\left(k+\frac32\right)(W+c-\nu)\ln W 
\nonumber\\
&+&\frac{1}{2}\left(\ln W-\frac12\right)W^{2} 
+(c-\nu)\,(\ln W-1)\,W
+{\cal G}\, .                     \label{b21c}
\eea
Here
\bea                           \label{b21d}
{\cal G}=
\int_0^\infty\frac{dt}
{{\rm e}^{2\pi t}-1}\, \{
t\,(6k(k+3)+2c+9-2t^2)\ln{\cal A} \nonumber\\
+(4k^3+18k^2+(18+4c)\,k+6c-6\,(2k+3)\,t^2)\Psi\}
\eea
with
\be                          \label{b21e}
{\cal A}=t^4+(2W-4\nu+9)\,t^2+W^2\, ,\ \ \ \
\Psi=\arctan\frac{(2k+3)\,t}{W-t^2}\, .
\ee
Evaluating the integral numerically, the two values 
used in the main text are 
\be                          \label{b21g}
Q'(2,0,-4|0)=3.72344\, , \ \ \ \ 
Q'(2,-4,0|0)=6.65246\, .
\ee
Finally, for the function 
$\zeta(s)=3^s(\frac{5}{12}Q(2,0,-4|s)-\frac14 Q(2,-4,0|s))$
used in Eq.(\ref{a68}) in the main text one obtains with 
the 
help of (\ref{b21b}) and (\ref{b21g})
\be
\zeta(0)=\frac{509}{90}\, ,\ \ \ \ 
\zeta'(0)\equiv\Gamm_1=6.10158\, .
\ee

\subsection{Computation of $Z(k,\nu|s)$}
Let us not turn to our main task -- the evaluation of the 
double-sum function $Z(k,\nu|s)$, which has been defined 
for large values of $\Re(s)$ by (\ref{A1}). 
The idea is to express it in terms of the single-sum 
function $\zeta(k,\nu|s)$. 

It follows from the definitions (\ref{A1}) and (\ref{A2}) that
\be					\label{b13}
Z(k,\nu|s)=\sum_{n=k}^\infty\,(2n+1)
 \zeta(k,\nu+(2n+1)^2|s)\, .
\ee
Here we can use the integral representation (\ref{b5})
for $\zeta(k,\nu+(2n+1)^2|s)$. Indeed, if $\nu$ is real
and $(2k+1)^2+\nu>0$ then the same remains true
upon replacement $\nu\to\nu+(2n+1)^2$, and the formula 
(\ref{b5}) therefore applies.  Now, 
replacing in (\ref{b5})  $\nu$ by $\nu+(2n+1)^2$ and
assuming for a moment that $\Re(s)$ is large and positive,
the integral in (\ref{b5}) converges uniformly with respect to 
$n$ for 
$n\to\infty$. This allows us, upon insertion of (\ref{b5})
into (\ref{b13}), to interchange the orders of summation
and integration. The result then can be extended to any $s$
by analytic continuation. This gives
\bea					\label{b13a}
Z(k,\nu|s)=
\left(k+\frac12\right)\sum_{n=k}^\infty
\frac{2n+1}{\{(2k+1)^2+\nu+(2n+1)^2\}^{s}}+ \nonumber \\
+\frac{1}{4(s-1)}
\sum_{n=k}^\infty
\frac{2n+1}{\{(2k+1)^2+\nu+(2n+1)^2\}^{s-1}} \nonumber 
\\
+\int_0^\infty\frac{id\tau}
{{\rm e}^{2\pi \tau}-1}
\left(
(2k+1+2i\tau)
\sum_{n=k}^\infty\frac{2n+1}{\left\{(2k+1+2i\tau)^2
+\nu+(2n+1)^2\right\}^s}
\right.\nonumber \\
\left.
-(2k+1-2i\tau)
\sum_{n=k}^\infty\frac{2n+1}{\left\{(2k+1-2i\tau)^2+
\nu+(2n+1)^2\right\}^s}
\right).\ \ \ \ \
\eea
One can see that all sums here are
exactly the same as  in the definition of 
$\zeta(k,\nu|s)$ in (\ref{A2}) -- up to the replacements
$\nu\to\nu+(2k+1)^2$ and 
$\nu\to\nu(\tau)\equiv\nu+(2k+1+2i\tau)^2$. 
Since the definition in (\ref{A2}) makes sense
for arbitrary values of $\nu$
(the series always converges for $\Re(s)$ big enough), 
we can express the sums in
(\ref{b13a}) in terms of $\zeta(k,\nu+(2k+1)^2|s)$ 
and $\zeta(k,\nu(\tau)|s)$. This leads to the   
the following formula:   
\bea                    \label{b14}
&&Z(k,\nu|s)=\left(k+\frac12\right)\zeta(k,\nu+(2k+1)^2|s)  \\
&&+\frac{1}{4(s-1)}\zeta(k,\nu+(2k+1)^2|s-1) 
+\int_0^\infty\frac{i\,d\tau}
{{\rm e}^{2\pi \tau}-1}\, \{{\cal F}(\tau)-{\cal F}(-\tau)\}\, 
,\nonumber
\eea
with ${\cal F}(\tau)=(2k+1+2i\tau)\,\zeta(k,\nu(\tau)|s)$. 
In this formula 
the first two terms on the right 
are determined 
by the integral representation (\ref{b5}) for arbitrary $s$. 
We are left with computing the remaining integral over $\tau$. 
The problem here is that the parameter
$\nu(\tau)$ is complex, and for this reason 
we cannot directly apply the 
integral representation (\ref{b5}) 
to compute $\zeta(k,\nu(\tau)|s)$.  

Let us recall that the formula (\ref{b5}) was derived assuming 
that 
the function $f(z)$ in Eq.(\ref{b4}) had no poles 
for  $\Re(z)>k$. This 
allowed us to rotate the integration contour
as shown in the left part of 
Fig.\ref{FIG-CONTOUR} without intersecting
singularities. Let us now replace $\nu$ by 
$\nu(\tau)\equiv\nu+(2k+1+2i\tau)^2$.
As a result, $f(z)$ in Eq.(\ref{b4}) is replaced by 
\be					\label{b15}
f(z)=\frac{2z+1}
{\left\{(2z+1)^2+\nu(\tau)\right\}^s}
=\frac{2z+1}
{\left\{4(z-z_{+}(\tau))(z-z_{-}(\tau))\right\}^s}\, ,
\ee 
with $z_{\pm}(\tau)=\frac12(-1\pm i\sqrt{\nu(\tau)})$. 
For $\tau=0$ one has $\Re(z_{\pm}(0))=-\frac12$. 
As $\tau$ increases, the point $z_{+}(\tau)$ moves to the right
in the complex plane (while $z_{-}(\tau)$ moves to the left), 
but as long as
$\Re(z_{+}(\tau))<k$ one can still use the formula (\ref{b5}).   
However, for large enough values of $\tau$
the pole at $z=z_{+}(\tau)$ enters the region of interest, 
that is the part of the complex plane with $\Re(z)>k$, 
and we can no longer use the formula (\ref{b5}). 

To tackle the problem we notice that the pole of $f(z)$ at
$z=z_{+}(\tau)$ is a branching point, and one can choose 
the cut in the complex plane as shown in the right part of
Fig.\ref{FIG-CONTOUR}. 
We then repeat the steps leading to the Abel-Plan formula
and the additional problem we encounter is the following:
when we rotate the integration contour as we did before,  
the contour will wrap
around the cut as shown in Fig.\ref{FIG-CONTOUR}. 
The resulting contour will then consist of two disconnected 
pieces. 
The first piece will be the same as the old contour 
$\bar{C}_1+C_2+\bar{C}_3$ 
(see Fig.\ref{FIG-CONTOUR}).  
The second piece is the 
contour $\tilde{C}$
wrapping around the cut. 
Integrating around
such a combined contour, 
the result will consist of two parts,
\be                          \label{b15a}
\zeta(k,\nu(\tau)|s)=\zeta_{\rm old}(k,\nu(\tau)|s)
+\theta(\tau-\tau_\ast)\int_{\tilde{C}}\frac{f(z)}{{\rm e}^{2\pi 
iz}-1}dz\, .
\ee
Here the first term on the right, 
$\zeta_{\rm old}(k,\nu(\tau)|s)$, is the function given by the 
previous expression in (\ref{b5}) with $\nu$ being replaced by 
$\nu(\tau)$. 
The second term, with $f(z)$ given by (\ref{b15}) 
and the contour $\tilde{C}$ as shown in the right part of
Fig.\ref{FIG-CONTOUR}, 
is the contribution of the cut. The step function 
$\theta(\tau-\tau_\ast)$
reflects
the fact that the cut contributes  
only for large enough $\tau$ 
when the pole enters the region $\Re(z)>k$. 
Here $\theta(x)=0$ for $x<0$ and $\theta(x)=1$ for $x\geq 0$,
and $\Re(z_{+}(\tau_\ast))=k$.   

The representation (\ref{b15a}) 
applies for all values of $s$ and for any $\tau>0$. 
Similarly,
one can obtain $\zeta(k,\nu(-\tau)|s)$  (the cut then
resides in the upper half-plane). As a result, the function
${\cal F}(\tau)-{\cal F}(-\tau)$ in the integrand in 
Eq.(\ref{b14})
is defined for any $\tau>0$, and the integral converges due to 
the damping 
exponential factor. This finally gives $Z(k,\nu|s)$ for any $s$. 

Let us first check our result by
computing $Z(k,\nu|0)$. For $s=0$ the function
$f(z)$ has no poles and the contribution of the cut
vanishes.  
The remaining integrals then can be easily computed, which gives 
for $Z(k,\nu|0)$ exactly the same expression as in 
Eq.(\ref{A19}).

Let us now compute $Z'(k,\nu|0)$. Since all
 integrals in (\ref{b14}),(\ref{b15}) 
converge uniformly with respect to $s$ (at least for 
$|s|<\infty$),
we can differentiate the integrands with respect to $s$ and then 
set $s=0$. 
The result can be represented in the following form:
\bea                        \label{b16}
Z'(k,\nu|0)={\cal H}+
2\int_0^\infty\frac{dt}
{{\rm e}^{2\pi t}-1}\,{\cal G}(t) \nonumber \\
+\int_0^\infty \frac{d\tau}{{\rm e}^{2\pi \tau}-1}
\int_0^\infty
\frac{d t}{{\rm e}^{2\pi t}-1}\,
{\cal W}(\tau,t)+{\cal S}. 
\eea
Here 
\bea                         \label{b17}
&&{\cal H}=\left(-2k^4+(2-\frac{\nu}{2})\,k^2+k+\frac{1}{32}\,
(4+4\nu-\nu^2)
\right)\ln(\,2\,(2k+1)^2+\nu)  \nonumber  \\
&&+3k^4+2k^3+\frac34\,(\nu-2)\,k^2
+\frac14\,(\nu-6)\,k+\frac{1}{64}\,(3\nu^2-4\nu-20)\, .\ \ \ \
\eea
In addition, 
\bea                          \label{b18}
{\cal G}(t)&=&\frac{t}{2}\,(4t^2-16k^2-12k-2-\nu)\ln P \\
&+&(2k+1)(6t^2-2k\,(2k+1)-\frac{\nu}{2})\,\Phi\, ,  \nonumber
\eea
where we have used 
\bea                           \label{b19}
P&=&\nu^2+(4(2k+1)^2-8t^2)\,\nu+4(2k+1)^4+16t^4\, , \nonumber \\
\Phi&=&{\rm Phase}[(2k+1)^2+\nu/2-2t^2+i\,2(2k+1)t]\, , 
\eea
and $-\pi<$Phase$[x+iy]\leq \pi$ is the phase of the complex 
number. 
Next, 
\bea                             \label{b20}
{\cal W}(\tau,t)&=&
\{((2k+1)^2-4\,t\,\tau)
\ln Q   \\
&+&4\,(t-\tau)(2k+1)\Psi\}-
\{(t,\tau)\leftrightarrow (t,-\tau)\} 
\eea
with 
\bea                              \label{b21}
Q&=&
16\,(t^2+\tau^2)^2  
+(\nu+2)^2-8\nu\,(t^2+\tau^2)
+128\,(k^2+k+\frac14)\,t\,\tau \nonumber \\
&+&16\,k(k+1)\,\nu+32\,k(k+1)(2k^2+2k+1),  \nonumber \\
\Psi&=&{\rm 
Phase}[(2k+1)^2+\nu/2-2\,(t^2+\tau^2)+i\,2\,(2k+1)(t-\tau)]\, .
\eea
Finally, the contribution of the cut is
\be                                \label{b22}
{\cal S}=4\pi\int_{\tau_\ast}^\infty\frac{d\tau}{{\rm 
e}^{2\pi\tau}-1}
\int_0^\infty \Im\,\frac{(2k+1-2i\tau)(2z(\tau)+1+2it)}
{{\rm e}^{2\pi(t-iz(\tau))}-1}\,dt \, ,
\ee
where 
$z(\tau)=-\frac12+\sqrt{4\tau^2-(2k+1)^2-\nu+4i(2k+1)\tau}$, 
and $\Re(z(\tau_\ast))=k$. 

We now use the formulas above in order to evaluate
$Z'(2,-10|0)$, which value is needed in the main text.
Setting $k=2$ and $\nu=-10$ we obtain for the first term on the 
right  in (\ref{b16}) ${\cal H}=1.9445$. The second term, 
containing the integral over $t$, is evaluated numerically to give 
$-19.9469$. The numerical value of the 
term containing the double integral is $-0.1294$. 
As for the last term, ${\cal S}$, it  is exponentially small
and is of the order of $10^{-12}$. 
This is because, as one can see from (\ref{b22}), the value
of ${\cal S}$ is suppressed by the factor of 
$\exp\{-2\pi(\tau_\ast+\Im(z(\tau_\ast))\}=\exp\{-4\pi\sqrt{5}\}$.
 
Summing everything up, we obtain
\be                 \label{ZETA}
\Gamm\equiv Z'(2,-10|0)=-18.3118\, .
\ee
 

\end{document}